\documentclass{aa}  

\usepackage{graphicx}
\usepackage{txfonts}
\usepackage{graphicx}
\usepackage{amsmath}
\numberwithin{equation}{section} %
\usepackage{mathtools}
\usepackage{commath}
\usepackage{siunitx}
\usepackage{times}
\usepackage{upgreek}
\usepackage{natbib}
\usepackage{paralist}
\usepackage{hyperref}
\usepackage[dvipsnames]{xcolor}

\usepackage{bm}
\defcitealias{DLMF}{DLMF}
\newcommand{\dlmf}[1]{(\citetalias[\S #1]{DLMF})}

\newcommand{\refedit}[1]{#1} %

\newcommand{\properbinom}[2]{\left(\!{{#1}\atop#2}\!\right)}
\newcommand{\expe}{\mathrm{e}}
\newcommand{\imagi}{\mathrm{i}}

\newcommand{\Beta}{\mathrm{B}}

\newcommand{\rs}{\ensuremath{r_s}}
\renewcommand\vec{\mathbf}
\newcommand{\intddd}{\ensuremath{\int \! d^{3\,}\!\vec{r}}}
\newcommand{\intdd}{\ensuremath{\int \! d^{2}\vec{R}}}
\newcommand{\intsphere}{\ensuremath{\int \! d^{2}\unitvector}}
\newcommand{\doubleint}{\ensuremath{\int d^\delta \vec{r}
                           \int d^\delta\vec{r}^\prime}}
\newcommand{\singleint}{\ensuremath{\int d^\delta \vec{r}^\prime}}
\newcommand{\unitvector}{\hat{\vec r}}
\newcommand{\cunitvector}{\hat{\vec R}}
\newcommand{\radiallaplacian}{\nabla^2_l}
\newcommand{\grad}{\bm{\nabla}}
\newcommand{\dop}{\ensuremath{\mathcal{D}}}
\newcommand{\dopp}{\ensuremath{\mathcal{A}}}
\newcommand{\rawpoly}{\ensuremath{P}}
\newcommand{\poly}{\ensuremath{p}}

\newcommand{\MellinTransform}[3]{\mathcal{M}_{#2}\!
                                 \left\{#1\right\}\!\!\left(#3\right)}
\newcommand{\InverseMellinTransform}[3]{\mathcal{M}^{-\!1}_{#2}\!\!
                                 \left\{#1\right\}\!\!\left(#3\right)}

\newcommand{\hyperg}[2]{\ensuremath{{}_{#1}{F}_{#2}}}
\newcommand{\hypergeom}[5]{\ensuremath{ {}_{#1} F_{#2}\!\left(\left.\!%
     \begin{matrix}
       #3 \\
       #4
     \end{matrix}%
   \: \!\right|\! #5 \!\right)}}

\begin{document}

\title{A general basis set algorithm for galactic haloes and discs}

\author{E. J. Lilley
  \inst{1}
  \and
  G. van de Ven\inst{2}
}

\institute{
  Department of Astrophysics, University of Vienna, T\"urkenschanzstra{\ss}e 17, 1180 Vienna, Austria\\
  \email{edward.lilley@univie.ac.at}
  \and
  \email{glenn.vandeven@univie.ac.at}  
}

\date{Received xxx; accepted yyy}

\abstract { We present a unified approach to (bi-)orthogonal basis
  sets for gravitating systems.
Central to our discussion is the notion of mutual gravitational
  energy, which gives rise to the \emph{self-energy inner product} on
  mass densities. We consider a first-order differential operator that
  is self-adjoint with respect to this inner product, and prove a
  general theorem that gives the conditions under which a
  (bi-)orthogonal basis set arises by repeated application of this
  differential operator.
  We then show that these conditions are fulfilled by all the families
  of analytical basis sets \refedit{with infinite extent that have
    been} discovered to date.
  The new theoretical framework turns out to be closely connected to
  Fourier-Mellin transforms, and it is a powerful tool for
  constructing general basis sets. We demonstrate this by deriving a
  basis set for the isochrone model and demonstrating its numerical
  reliability by reproducing a known result concerning unstable radial
  modes.  }


   \maketitle

\section{Introduction}
\label{sec:intro}

Orthogonal basis sets play a key role in the efficient calculation of
the gravitational potential of perturbed, isolated mass
distributions. They also have great value for investigating the
stability of dynamical models for galaxies. Both these topics have
attracted renewed interest recently in light of the mounting
observational evidence that the Milky Way and other galaxies are not
as symmetric in shape as assumed previously \citep{Ve13,La10}, and
moreover may not be in exact dynamical equilibrium \refedit{\citep{Er21,Pe21}}.

A small sample of recent applications of basis sets includes:
efficiently reconstructing individual trajectories in time-varying
snapshots of $N$-body simulations of dark matter haloes
\refedit{\citep{Lo11,Sa20,Pe22b}}; flexible non-parametric models for the Milky Way
\citep{Ga21}; and a wide variety of perturbation calculations
\citep{Ha18,Fouvry2022}.

The development of these so-called `biorthogonal' basis sets begins
with \citet{CB72,CB73}, who introduced two remarkable analytical sets
of potential-density pairs based on the \citet{Kuz56} disc and
\citet{Plu1911} sphere respectively. These mathemtical discoveries
(along with some later results discussed below), while fortunate, are
limited. It has long been recognised that to make best use of the
basis set technique, one would prefer a complete freedom in choice of
zeroth-order (as well as underlying coordinate system and geometry),
while making minimal sacrifice of computational efficiency.

To this end, there are basically three possible directions of
generalisation. One might hope to have the good fortune of finding
other `analytical' basis sets, taking some known model as the
zeroth-order potential-density pair and then hoping that by some
ingenious change-of-variables or integral transform a set of
orthogonal higher-order functions can be written down. This approach
is limited but has provided a handful of further results in both
spherical polar coordinates \citep{HO92,Zh96,Ra09,LSEE,LSE} and for
infinitesimally thin discs \citep{Ka76,Qi93}. Generally speaking, for
both spheres and thin discs, basis sets exist for some double
power-laws and for certain types of exponential distributions of mass.

Secondly, one could posit an arbitrary sequence of non-orthogonal
potential-density pairs, and from them derive an orthogonal set using
the Gram-Schmidt algorithm. This is the approach of
\citet{Sa93,Ro96}. The downsides are the large number of expensive
numerical integrations required to compute the required inner
products, the numerical instability inherent to the Gram-Schmidt
process, and the uncertain completeness or convergence properties of
the resulting orthogonal basis.

Lastly, the strategy devised by \citet{We99,Pe22} generalises
\citet{CB73}'s original result directly by noticing that the
potential-density relation takes the form of a Sturm-Liouville
eigenfunction equation with a certain weight function; by choosing a
different weight function and using a numerical Sturm-Liouville
solver, a different set of eigenfunctions (and hence basis set) can be
found. This approach has the upside that certain guarantees about
completeness and convergence can be made, but the downside that the
resulting eigenfunctions must be tabulated numerically on a coordinate
grid.

In this paper we describe a different generalisation of
Clutton-Brock's original results -- we jettison the eigenfunction
equation but retain a three-term recurrence relation.

Essentially our approach is motivated by the observation that the
extant basis sets\footnote{Of analytical form and infinite extent.} so
far described in the literature admit the curious property of
\emph{tridiagonality} with respect to a radial derivative
operator. That is, for a given density basis function
$\rho_n(\vec{r})$ (suppressing the angular indices and coordinates),
the following holds,
\begin{equation}\label{eq:simple_tridiag}
  r \frac{\partial \rho_n}{\partial r} = a_n \rho_{n-1} + b_n \rho_n + c_n \rho_{n+1},
\end{equation}
where $a_n$, $b_n$, $c_n$ are constants. This may seem to be merely a
curiosity, but upon further reflection it motivates a far-reaching
generalisation: armed with just knowledge of an arbitrary (smooth)
zeroth-order basis element, the tridiagonality property
\eqref{eq:simple_tridiag} allows us to build up an entire ladder of
basis elements recursively, using just one additional integral per
recursive step. The resulting basis elements are linear combinations
of derivatives of the zeroth order and hence require no further
interpolation. Along these lines, in Sec.~\ref{sec:new} we present an
algorithm to generate general basis sets from arbitrary zeroth-order
potential-density pairs.

Underlying this main result is a link to the general theory of
orthogonal polynomials, which motivates us to claim completeness of
the resulting basis sets. This theoretical background is discussed in
Sec.~\ref{sec:theory}, where we introduce the Fourier-Mellin
transform, and show a correspondence between tridiagonal orthogonal
basis sets and orthogonal polynomials in the transformed space. Key to
this link is the notion of the gravitational self-energy inner
product, and an operator ($\dop$) that is self-adjoint with respect to
it.

The new approach was in part first suggested implicitly by
\citet{Ka76}, who introduced the Fourier-Mellin transform (but not
naming it as such) in the case of \refedit{thin
  discs\footnote{\refedit{Throughout this work we will use `thin disc' to refer
    to an idealised infinitesimally thin disc; basis sets for discs
    with nonzero thickness are out of scope, albeit an important
    future direction of research.}}}, but nevertheless only used it to
rederive the \cite{CB72} basis set. Those results are partially
repeated (using our updated notation) in Sec.~\ref{sec:disc}, where we
show that a formalism equivalent to the spherical case exists for
\refedit{thin discs} in cylindrical polar coordinates, along with a
similar self-adjoint operator ($\dopp$).

As further motivation for our new algorithm, in
Sec.~\ref{sec:existing} we demonstrate concretely how the formalism
applies to some existing basis sets in the literature. Specifically we
show that the two major families of basis sets -- corresponding to
double-power laws in the spherical \citep{LSE} and \refedit{thin disc
  \citep{Qi93} scenarios} (along with their various limiting forms) --
\refedit{both} possess the tridiagonality property, and hence each
admit a representation in terms of a polynomial in $\dop$ or $\dopp$
respectively.

In Sec.~\ref{sec:numerical} we return to the general algorithm
described in Sec.~\ref{sec:new}, and discuss the numerical and
computational issues that arise when trying to implement it in
practice. In particular it is necessary to find a fast, stable method
to evaluate the requisite numerical integrals. This is most easily
accomplished using Gauss-Laguerre quadrature in the transformed
(Fourier-Mellin) space, first computing the underlying system of
orthogonal polynomials. The recommended procedure is illustrated with
the case of the isochrone model, which we use in
Sec.~\ref{sec:application} to recover a known result about unstable
radial modes.

Finally in Sec.~\ref{sec:discussion} we discuss some of the geometric
ideas underlying the new formalism. We outline how our results might
be extended to other geometries or coordinate systems relevant to the
study of realistic galaxies, and give an outlook on future work to be
done in the area.

\section{Description of algorithm}
\label{sec:new}

First we make some new definitions as well as recapitulating the
standard terminology. We define the \emph{self-energy inner product}
$\langle\cdot, \cdot\rangle$ on mass densities
\begin{equation}
\label{eq:innerproduct}
\langle\rho_1, \rho_2\rangle = \intddd \! \intddd^\prime \:
\frac{\rho_1(\vec{r}) \overline{\rho_2(\vec{r}^\prime)}}
     {\lVert \vec{r} - \vec{r}^\prime\rVert}.
\end{equation}
This is sometimes referred to as the mutual gravitational potential
energy of $\rho_1$ with respect to $\rho_2$. Of course, the total
self-energy is just
$\lVert \rho \rVert^2 = \langle\rho, \rho\rangle$,
which here must clearly always be real and positive (although the
normal convention is for this quantity to be negative, the overall
choice of sign is irrelevant for our purposes).  It is important that
\eqref{eq:innerproduct} obeys the standard properties of a inner
product: linear in its first and conjugate linear in its second
argument. Generically we allow mass densities to be complex-valued, as
it eases some of the following derivations; however the entire
formalism (necessarily) also works in the case of purely real mass
densities. We are also not limited to densities with finite total
mass, only finite total self-energy\footnote{Many popular mass laws
  have infinite mass but finite self-energy, e.g.~the NFW model
  \citep{NFW}.}. Finally we note that if we have a solution to
Poisson's equation for $\rho_1$ and $\rho_2$, finding their
gravitational potentials to be $\Phi_1$ and $\Phi_2$, then (using
Green's identities) we can rewrite the inner product
\eqref{eq:innerproduct} as
\begin{equation}
  \langle\rho_1, \rho_2\rangle = \frac{1}{4\pi}
   \intddd \: \grad \Phi_1 \cdot \overline{\grad \Phi_2},
\end{equation}
or alternatively as
\begin{equation}
\langle\rho_1, \rho_2\rangle = -\!\intddd \: \Phi_1 \overline{\rho_2}.
\end{equation}
We set the gravitational constant $G = 1$ throughout.  Now we
introduce both spherical polar coordinates $(r,\varphi,\vartheta)$ and
cylindrical polar coordinates $(R,\varphi,z)$, the latter being used
here only in the situation where the mass density is confined to
\refedit{a thin disc} aligned with the $z$-axis. We define two
important operators,
\begin{equation}\label{eq:zeta_def}
\dop = \imagi \left(r\partial_r + \frac{5}{2}\right)
\end{equation}
and
\begin{equation}\label{eq:xi_def}
\dopp = \imagi \left( R\partial_R + \frac{3}{2} \right).
\end{equation}
These have the important property of being self-adjoint with
respect to the inner product \eqref{eq:innerproduct} (see
App.~\ref{sec:zeta_selfadjoint} for a proof), i.e.
\begin{equation}\label{eq:zeta_selfadjoint}
\langle \dop f, g \rangle = \langle f, \dop g \rangle,
\end{equation}
and (when $f$ and $g$ are \refedit{thin discs})
\begin{equation}\label{eq:xi_selfadjoint}
\langle \dopp f, g \rangle = \langle f, \dopp g \rangle.
\end{equation}

Our standard notation for basis sets is as follows. We denote by
$\{\rho_{nlm}\}$ a complete basis for the set of smooth mass densities
satisfying:
\begin{align}\label{eq:valid_rho}
  \lVert \rho \rVert^2 &< \infty \qquad \text{(finite self-energy)}, \\ \nonumber
  \rho(\vec{r}) &= 0 \: \text{at only isolated points} \: \vec{r} \qquad \text{(infinite extent)}.
\end{align}
The set $\{\rho_{nlm}\}$ is assumed orthogonal with respect to
\eqref{eq:innerproduct},
\begin{equation}
\label{eq:spherical_basis_orth}
\langle\rho_{nlm}, \rho_{n^\prime l^\prime m^\prime}\rangle
 = N_{nlm} \delta_{nlm}^{n^\prime l^\prime m^\prime}, \qquad N_{nlm} = -K_{nl} N_{nl}.
\end{equation}
These basis functions are the product of radial and angular components,
\begin{align}\label{eq:spherical_basis}
\Phi_{nlm}(\vec{r}) &= \Phi_{nl}(r) \: Y_{lm}(\unitvector), \\ \nonumber
\rho_{nlm}(\vec{r}) &= K_{nl} \: \rho_{nl}(r) \: Y_{lm}(\unitvector),
\end{align}
which satisfy
\begin{align}\label{eq:spherical_basis_laplacian}
\nabla^2\Phi_{nlm} &= 4 \pi \rho_{nlm}, \\ \nonumber
\radiallaplacian\Phi_{nl} &= 4 \pi K_{nl} \rho_{nl}.
\end{align}
where $K_{nl}$ are constants factored out of $\rho_{nl}$ just to
simplify the expressions; and $\radiallaplacian$ is the radial part of
the Laplacian when operating on
$(\text{radial functions})\times(\text{a spherical harmonic of
  order}\:l)$:
\begin{equation}\label{eq:radial_laplacian_def}
\radiallaplacian = r^{-2} \frac{d}{dr} \left(r^2 \frac{d}{dr}\right) - \frac{l(l+1)}{r^2}.
\end{equation}
The purely radial functions $\rho_{nl}(r)$ and $\Phi_{nl}(r)$ are
real-valued, and satisfy a `bi-orthogonality relation'
\begin{equation}
\int_0^\infty dr \: r^2 \: \Phi_{nl} \: \rho_{nl} = N_{nl} \delta_{n n^\prime}.
\end{equation}
For this reason such basis sets are traditionally referred to as
\emph{bi-orthogonal}. Note that we take $Y_{lm}$ throughout to be a
unit-normalised (complex) spherical harmonic. If non-orthonormal
spherical harmonics are employed then $N_{nlm}$ must contain the
appropriate factor that normalises them. The radial functions
$\Phi_{nl}$ and $\rho_{nl}$ are typically functions of the quantity
$r/\rs$, where \rs\ is some `scalelength' with units of length; we
will generally use $\rs = 1$ implicitly\footnote{Explicit length units
  can be reintroduced by writing $\Phi_{nl}(r/\rs)$ and
  $\rho_{nl}(r/\rs)$ and then adding the correct number of powers of
  \rs\ in whatever expression they are used. Note that such
  \rs-dependency cancels out in the operators $\dop$ and $\dopp$.}.

An analogous notational convention is used throughout for the case of
a \refedit{thin disc}. We write $\{\sigma_{nm}\}$ to represent a complete
basis, where
\begin{equation}
  \nabla^2\left( \psi_{nm}(R,z) \: \expe^{\imagi m \varphi} \right) = \sigma_{nm}(R,z) = \sigma_{nm}(R) \: \expe^{\imagi m \varphi} \: \delta(z).
\end{equation}
In an abuse of notation we suppress the $z$-dependence and elide the
quantities which have subscript $nm$, writing the potential in the
disc plane as
\begin{equation}
  \psi_{nm}(\vec{R}) = \psi_{nm}(R) \: \expe^{\imagi m \varphi}.
\end{equation}

We now describe a natural method for deriving basis sets with
\emph{any} smooth analytical zeroth-order element. We will focus on
the spherical case, and afterwards describe the (slight) changes
required in the \refedit{thin disc} case.

The first step is to choose a suitable zeroth-order potential, which
we denote $\Phi(r)$. This can be chosen according to the problem at
hand, the only requirements being that it must be a smooth
spherically-symmetric function of $r$, \refedit{and the
  potential-density pair must have finite total gravitational
  self-energy}. Starting from $\Phi$ we must then invent a function
$\Phi_{0l}(r)$ that provides the zeroth radial order for the higher
multipoles indexed by $l$. This function must satisfy two boundary
conditions\footnote{\refedit{For models with infinite enclosed mass
    the potential can contain an additional factor of $\log{r}$ as
    $r \to \infty$.}}: $\Phi_{0l} \sim r^l$ as $r \to 0$ and
$\Phi_{0l} \sim r^{-l-1}$ as $r \to \infty$. One way to achieve this
is to take
\begin{equation}\label{eq:phi0l_def}
  \Phi_{0l}(r) = r^l \left[ \Phi(r) \right]^{2l+1},
\end{equation}
but any choice with the correct asymptotic behaviour will do just as
well\footnote{Other choices may be preferable from an analytical point
  of view, for example $\Phi_{0l} = r^l \Phi\!\left(r^{2l+1}\right)$
  or $\Phi_{0l} = r^l \Phi(r)/(1+r)^{2l}$, the latter suggested by
  \citet{Sa93}.} Once $\Phi_{0l}$ is chosen, the corresponding density
multipoles $\rho_{0l}$ are fully determined by
\begin{equation}
\radiallaplacian\Phi_{0l} = 4\pi K_{0l}\rho_{0l},
\end{equation}
where $K_{0l}$ is an arbitrary constant chosen to simplify the
algebra.

The defining relation for the basis set with zeroth order $\rho_{0l}$
is the differential-recurrence relation,
\begin{equation}\label{eq:rhonl_recurrence}
  \rho_{n+1,l} = \left(r\partial_r + \frac{5}{2}\right)\rho_{nl} + \beta_{nl}\rho_{n-1,l},
\end{equation}
with initial conditions $\rho_{-1,l} = 0$, and where $\beta_{nl}$ are
some (as yet undetermined) constants. Note that the operator applied
to the $\rho_{nl}$ term on the RHS is equal to $-\imagi\dop$
\eqref{eq:zeta_def}. We can immediately write down a similar
recurrence for the potential elements,
\begin{equation}\label{eq:Phinl_recurrence}
  \Phi_{n+1,l} = \left(r\partial_r + \frac{1}{2}\right)\Phi_{nl} + \beta_{nl}\Phi_{n-1,l},
\end{equation}
due to the commutation relation between $\dop$ and the radial
Laplacian $\radiallaplacian$ (see App.~\ref{sec:commutator_proof}). By
taking the inner product of \eqref{eq:rhonl_recurrence} with both
$\rho_{n+1,l}$ and $\rho_{n-1,l}$, and exploiting the self-adjointness
property \eqref{eq:zeta_selfadjoint}, we find that the constants
$\beta_{nl}$ are given by
\begin{equation}\label{eq:betanl_def}
\beta_{nl} = \frac{\lVert \rho_{nl} \rVert^2}{\lVert \rho_{n-1,l} \rVert^2}.
\end{equation}
This is just the ratio of the gravitational self-energy of the $n$th
and $(n-1)$th basis elements. Because the RHS of
\eqref{eq:rhonl_recurrence} depends only on the $n$th and lower
elements, we can now build up the entire sequence of basis elements by
alternating applications of \eqref{eq:rhonl_recurrence} and
\eqref{eq:betanl_def}.

This deceptively simple algorithm leaves some unresolved issues:
\begin{inparaenum}
\item Are these basis sets truly complete?
\item How do we deal with the differentiation required in \eqref{eq:rhonl_recurrence}?
\item Are the numerical integrals in \eqref{eq:betanl_def} stable?
\end{inparaenum}

We can give at least convincing heuristic answers to these
questions. The question of completeness we consider in the course of
the theoretical discussion in Sec.~\ref{sec:theorem}. The repeated
differentiation will in general require some form of \emph{symbolic}
or \emph{automatic} differentiation, which we discuss in
Sec.~\ref{sec:repeated_diff} -- unless the specific form of the
zeroth-order allows for a simplification. The question of numerically
calculating the recurrence coefficients $\beta_{nl}$ is thorny, and we
return to it in Sec.~\ref{sec:numerical} after developing in
Sec.~\ref{sec:theory} the theoretical machinery that links these basis
sets to the theory of general orthogonal polynomials.

Our resulting basis elements are linear combinations of the
higher-derivatives of the zeroth-order functions:
$\{\dop^n\rho_{0lm}\}$ in the case of the density, and
$\{\dop^n\Phi_{0lm}\}$ in the case of the potential. This means that,
given a closed-form zeroth-order, all higher elements are generated
through differentiation -- no numerical interpolation is required,
unlike \citet{We99}'s algorithm based on Sturm-Liouville
eigenfunctions. In fact, given a particular zeroth-order, a basis
computed via the Sturm-Liouville approach will \emph{not} in general
coincide with the basis set developed from our own algorithm, except
for certain special cases that are known to obey eigenfunction
equations (for example the \citet{Zh96} basis sets).

In addition, unlike \citet{Sa93}, we are able to avoid the brute force
approach of Gram Schmidt orthogonalisation (with complexity $O(n^2)$
in the number of inner products, and uncertain numerical
stability). This is due to the self-adjointness of the operator
$\dop$, which ensures that each basis set maps onto an underlying
orthogonal polynomial in Fourier-Mellin space, a mathematical
connection elaborated \refedit{upon} in Sec.~\ref{sec:theory}. Thus we
can reuse the large body of literature regarding the construction of
general orthogonal polynomials, the most important property being that
any set of orthogonal polynomials obeys a three-term recurrence
relation -- this relation is transferred over to the basis set,
manifesting as the differential-recurrence relation
\eqref{eq:rhonl_recurrence}.

Lastly we note that in the case of a \refedit{thin disc} the surface
densities $\sigma_{nm}$ have fundamental differential-recurrence
relation
\begin{equation}\label{eq:sigmanm_recurrence}
  \sigma_{n+1,m} = \left(R\partial_R + \frac{3}{2}\right)\sigma_{nm} + \beta_{nm}\sigma_{n-1,m},
\end{equation}
where the operator applied to $\sigma_{nm}$ on the RHS is now
$-\imagi\dopp$ \eqref{eq:xi_def}; but the algorithm is otherwise
identical to the spherical case. In both the spherical and
\refedit{thin disc} case the algorithm can be initialised by choosing
either the zeroth-order potential or the zeroth-order density; but
starting with a density may be more difficult in the \refedit{thin
  disc} case as analytical potential-density pairs are harder to come
by. \refedit{The required boundary conditions on $\psi_{0m}$ (with
  azimuthal index $m$ standing in for $l$) are unchanged, as is the
  requirement of smoothness and finite self-energy.}

\section{Theoretical background}
\label{sec:theory}
\subsection{Functional calculus of $\dop$ and the Fourier-Mellin transform}
\label{sec:fourier_mellin_intro}

Consider the eigenfunctions of $\dop$, which we denote
$\Psi_s$. These satisfy
\begin{equation}
\dop \Psi_s = s \Psi_s, \qquad s \in \mathbb{R}
\end{equation}
and have the form
\begin{equation}\label{eq:psi_s}
\Psi_s(r) = r^{-\imagi s - 5/2}.
\end{equation}
We combine this with a spherical harmonic to define the
$\dop$-\emph{eigenbasis}
\begin{equation}\label{eq:psi_slm}
\Psi_{slm}(\vec{r}) = \Psi_s(r) \: Y_{lm}(\unitvector).
\end{equation}
Now let $F(\vec{r})$ be a general mass density, and $F_{lm}(r)$ its
spherical multipole moments. Then the expansion coefficient of $F$ in
the $\dop$-eigenbasis is (see App.~\ref{sec:psi_inner_product} for
proof)
\begin{equation}\label{eq:fourier_mellin}
  \langle F, \Psi_{slm}\rangle = \frac{4\pi}{K_l(\imagi s)} \MellinTransform{F_{lm}(r)}{r}{5/2 + \imagi s},
\end{equation}
where $K_l(\imagi s)$ is defined through
\begin{equation}\label{eq:Kl}
K_l(s) = (l + 1/2 + s)(l + 1/2 - s),
\end{equation}
and $\mathcal{M}_r$ is the Mellin transform,
\begin{equation}\label{eq:mellin}
\MellinTransform{f(r)}{r}{s} = \int_0^\infty r^{s-1} \: f(r) \: dr.
\end{equation}
We will refer (with some precedent) to the combination
\eqref{eq:fourier_mellin} of taking multipole moments and a Mellin
transform as the three-dimensional \emph{Fourier-Mellin transform}. We
can re-express $F$ in terms of its
Fourier-Mellin expansion coefficients using the Mellin inversion
theorem (via an appropriate change of variable),
\begin{equation}
  F(\vec{r}) = \frac{1}{8\pi^2} \sum_{lm} \int_{-\infty}^\infty ds
  \: K_l(\imagi s) \: \Psi_{slm}(\vec{r}) \:
  \langle F, \Psi_{slm} \rangle.
\end{equation}
where the inverse Mellin transform $\mathcal{M}^{-1}$ is
\begin{equation}\label{eq:mellin_inversion}
  \InverseMellinTransform{g(s)}{s}{r} = \frac{1}{2\pi\imagi} \:
  \int_{c-\imagi \infty}^{c+\imagi\infty} r^{-s} \: g(s) \: ds
\end{equation}
for some constant $c$, the choice of which does not affect any of our
results. The mutual gravitational energy of two
general mass densities $F$ and $G$ can therefore be expressed as
\begin{equation}\label{eq:innerproduct_general}
  \langle F, G \rangle = \frac{1}{8\pi^2} \sum_{lm}
  \int_{-\infty}^{\infty} ds \: K_l(\imagi s) \: \langle F, \Psi_{slm} \rangle \langle \Psi_{slm}, G \rangle.
\end{equation}
Because $\dop$ is self-adjoint the spectral theorem applies, and we
can consider arbitrary bounded complex-valued functions of $\dop$. The
Fourier-Mellin transform can be viewed as the (unitary) map to the
space in which $\dop$ acts as a multiplication operator. In practice
though, we can limit ourselves to considering polynomials in $\dop$.

The formalism developed above also applies mutatis mutandis to the
\refedit{thin disc} case. The derivation is now mostly the same as
that found in \citet{Ka71,Ka76}, but we update his notation. Our
self-adjoint operator $\dopp$ has eigenfunctions $\Sigma_s$ satisfying
\begin{align}
  \dopp \Sigma_s &= s \Sigma_s, \\ \nonumber
  \Sigma_s(R) &= R^{-\imagi s - 3/2}, \\ \nonumber
  \Sigma_{s m}(\vec{R}) &= \Sigma_s(R) \: \expe^{\imagi m \varphi}.
\end{align}
The functions $\Sigma_{s m}(\vec{R})$ are Kalnajs' \emph{logarithmic
  spirals}\footnote{\refedit{See e.g.~\citet[Eq.~(14)]{Ka71}, set
    $u = \log{R}$ and relabel $\theta \to \varphi$ and
    $\alpha \to -s$. The RHS there is then proportional to our
    $\Sigma_{sm}(\vec{R})$, apart from a factor of $R^{-3/2}$.}}. For
a general razor-thin mass density $\sigma(\vec{R})$ we have the thin
disc version of the Fourier-Mellin transform (see
App.~\ref{sec:disc_fourier_mellin} for proof),
\begin{equation}
  \langle \sigma, \Sigma_{sm} \rangle = \frac{\pi}{K_m(\imagi s)} \MellinTransform{\sigma_m(R)}{R}{3/2+\imagi s},
\end{equation}
where $\sigma_m(R)$ are the cylindrical multipoles of
$\sigma(\vec{R})$, and
\begin{equation}
  K_m(\imagi s)=\left| \frac{\Gamma\!\left(\frac{m+3/2+\imagi s}{2}\right)}
  { \Gamma\!\left(\frac{m + 1/2 + \imagi s}{2}\right) } \right|^2.
\end{equation}

\subsection{Tridiagonality and polynomials}\label{sec:theorem}

Associated to each of our basis sets is a polynomial \refedit{we
  refer} to as the \emph{index-raising polynomial} -- depending on the
normalisation we write either $\poly_{nl}(s)$ or $\rawpoly_{nl}(s)$
(for a polynomial of degree $n$ in the variable $s$). The general
result proved below is that applying the $n$th-degree polynomial with
argument $\dop$ to the zeroth-order density element gives the
$n$th-order density element. It may help with interpretation to note
that these polynomials in a sense `live' in the Fourier-Mellin space
introduced in the previous section.

There are several related statements that one can make about a given
basis set and its associated index-raising polynomial:
\begin{enumerate}
\item The tridiagonality of the density basis functions
  $\{\rho_{nlm}\}$ with respect to the operator $\dop$;
\item The expressibility of each basis function in terms of a
  polynomial in $\dop$ applied to the lowest-order basis function,
  with these polynomials obeying a three-term recurrence relation;
\item The orthogonality of $\poly_{nl}(s)$
  with respect to a weight function $\omega_l(s)$ given in terms of
  the Mellin transform of $\rho_{0l}$;
\item The orthogonality of the basis functions $\{\rho_{nlm}\}$ with
  respect to the self-energy inner product
  $\langle\cdot,\cdot\rangle$.
\end{enumerate}
Below we show that the first and second statements are equivalent. We
also find that the third statement implies the fourth, and the second
and fourth together imply the third.  However, while it is easy to
show that the third statement implies the second, the converse is much
harder. Favard's theorem guarantees that a set of polynomials obeying
a three-term recurrence relation is orthogonal with respect to
\emph{some} measure, however this is a difficult computation and is
not what we actually want. In practice we want the freedom to specify
zeroth-order basis elements, not the recurrence coefficients
themselves.

Therefore, to construct an arbitrary basis set we \emph{impose} the
first and fourth statements. Then the second and third statements
(which provide the polynomials $\rawpoly_{nl}(s)$ or $\poly_{nl}(s)$)
are a useful representation of the underlying basis set, which we
exploit in order to solve numerical issues in the implementation
described in Sec.~\ref{sec:numerical}.

The idea of finding orthogonal polynomials from tridiagonal matrices
or operators is not new; in the finite-dimensional case the
corresponding matrix is called a \emph{Jacobi matrix}, and gives rise
to polynomials of discrete argument. Our work invokes the
infinite-dimensional case, in which a \emph{Jacobi operator} (here
$\dop$ or $\dopp$) operates on an infinite sequence of functions,
which we generally assume to be a complete orthogonal set that spans
the relevant function space. Such infinite-dimensional Jacobi
operators are studied in
\citet{Granovskii1986,Ismail2011,Dombrowski1985}, and our set-up
mimics the development given in the first paper, with the difference
that our $\dop$ and $\dopp$ are taken as given and do not arise from
any Lie algebraic considerations\footnote{The operators $\dop$ and
  $\dopp$ do in fact arise as the generators of symmetries of the
  self-energy inner product; see the discussion in
  Sec.~\ref{sec:discussion}.}.

As in the previous section, we give the main derivations in the case
of spherical \refedit{polar coordinates}; the thin disc case then
follows with little modification.

\subsubsection{Polynomials from tridiagonality}

We show that any set of densities $\{\rho_{nlm}\}$ that is
\emph{tridiagonal with respect to $\dop$} gives rise to an expression
for each $\rho_{nlm}$ in terms of an index-raising polynomial in
$\dop$, of the form
\begin{equation}\label{eq:pirho}
  \rho_{nlm} = \rawpoly_{nl}(\dop) \rho_{0lm}.
\end{equation}
By \emph{tridiagonality} we mean that the following expression holds,
\begin{equation}
\label{eq:tridiag}
\dop \rho_{nlm} = a_{nl} \rho_{n-1,lm}\!+\!b_{nl} \rho_{nlm}
                   \!+\! c_{nl} \rho_{n+1,lm},
\end{equation}
for some constants $a_{nl}$, $b_{nl}$ and $c_{nl}$. First, define
\begin{equation}
\chi_{nlm} = \dop^n \rho_{0lm}.
\end{equation}
From \eqref{eq:tridiag} there exists an expansion of $\chi_{nlm}$ of the form
\begin{equation}\label{eq:eta_rho_expansion}
\chi_{nlm} = \sum_{j=0}^n B_{nlj} \rho_{jlm}.
\end{equation}
Then by inverting $B_{njl}$ (interpreted as a matrix with respect to
the $nj$ indices) it is evidently possible to write an expansion for
$\rho_{nlm}$ of the form
\begin{equation}
\label{eq:rho_eta_expansion}
\rho_{nlm} = \sum_{j=0}^n A_{nlj} \: \chi_{jlm}.
\end{equation}
Now make the definition
\begin{equation}\label{eq:pi_def}
  \rawpoly_{nl}(s) = \frac{\langle\Psi_{slm},\rho_{nlm}\rangle}
          {\langle\Psi_{slm},\rho_{0lm}\rangle}.
\end{equation}
To prove that $\rawpoly_{nl}(s)$ is a polynomial, take the
Fourier-Mellin expansion
of \eqref{eq:rho_eta_expansion},
\begin{align}
\langle\Psi_{slm}, \rho_{nlm}\rangle &= \sum_{j=0}^n A_{njl}
  \langle\Psi_{slm},\chi_{jlm}\rangle \\ \nonumber
&= \sum_{j=0}^n A_{njl} \: s^j \: \langle\Psi_{slm}, \rho_{0lm}\rangle,
\end{align}
where the second equality uses the self-adjointness property
\eqref{eq:zeta_selfadjoint} as well as the definition of the
eigenbasis \eqref{eq:psi_slm}. Dividing through by
$\langle\Psi_{slm}, \rho_{0lm}\rangle$ then gives $\rawpoly_{nl}(s)$
as a polynomial in $s$ with (as yet undetermined) coefficients
$A_{njl}$. But from the definition \eqref{eq:rho_eta_expansion} we see
that $\rawpoly_{nl}(\dop)$ is just the operator expression for
$\rho_{nlm}$ that raises the radial index from $0$ to $n$, which is
\eqref{eq:pirho}.
To find the three-term recurrence relation for $\rawpoly_{nl}(s)$,
take the
Fourier-Mellin expansion of \eqref{eq:tridiag}, divide through by
$\langle\Psi_{slm}, \rho_{0lm}\rangle$, and rearrange, giving
\begin{equation}
\label{eq:pi_recurrence}
\rawpoly_{n+1,l}(s) = \frac{s - b_{nl}}{c_{nl}} \rawpoly_{nl}(s)
- \frac{a_{nl}}{c_{nl}} \rawpoly_{n-1,l}(s).
\end{equation}
For the converse statement, substituting $\dop$ for $s$ in the above
recurrence and left-applying to $\rho_{0lm}$ trivially recovers the
tridiagonality property (we must also take the initial conditions
$\rawpoly_{0l} = 1$ and $\rawpoly_{-1l} = 0$).

\subsubsection{Orthogonal polynomials}

From Favard's theorem we know that the $\rawpoly_{nl}(\dop)$ are a
system of orthogonal polynomials, as they satisfy a three-term
recurrence relation \eqref{eq:pi_recurrence}. However, in order to
actually construct the orthogonalising weight function, we first
assume that the underlying basis functions are already orthogonal. It
follows that
\begin{align}\label{eq:rho_poly_orth}
\langle \rho_{nlm},\rho_{n^\prime l^\prime m^\prime} \rangle
&= \delta_{ll^\prime} \delta_{mm^\prime}
\int_{-\infty}^\infty ds \: \omega_l(s) \: \rawpoly_{nl}(s)
                                                               \:\overline{\rawpoly_{n^\prime l}(s)} \\ \nonumber
  &\propto \delta_{nn^\prime} \delta_{ll^\prime} \delta_{mm^\prime},
\end{align}
where the (positive, real-valued) weight function $\omega_l(s)$ is
related to the zeroth-order density basis function $\rho_{0l}$ by
\begin{equation}
\label{eq:omega}
\omega_l(s) = \frac{2 K_{0l}^2}{K_l(\imagi s)} \left|
\MellinTransform{\rho_{0l}(r)}{r}{5/2 + \imagi s} \right|^2,
\end{equation}
the proof of which is in App.~\ref{sec:orth_proof}. The orthogonality
relation \eqref{eq:rho_poly_orth} works in both directions: if we
instead assume that $\rawpoly_{nl}(s)$ are orthogonal with respect to
(a given) $\omega_l(s)$, then the orthogonality of the $\rho_{nlm}$
follows.

In fact, $\rawpoly_{nl}(s)$ can be written in terms of purely real
polynomials $\poly_{nl}(s)$, which are also orthogonal with respect to
$\omega_l(s)$,
\begin{equation}
  \int_{-\infty}^\infty \omega_l(s) \: \poly_{nl}(s) \: \poly_{n^\prime l}(s) \: ds \propto \delta_{n n^\prime},
\end{equation}
with
\begin{equation}
  \rawpoly_{nl}(s) \propto \imagi^{-n} \: \poly_{nl}(s).
\end{equation}
It is often more convenient in applications to deal with these
real-valued polynomials. Without loss of generality (up to
normalisation) we can take the polynomials $\poly_{nl}(s)$ to be
monic\footnote{\emph{Monic} meaning that the term of highest-degree
  has coefficient $1$. Note that in Sec.~\ref{sec:existing} the
  polynomials are not necessarily in monic form.}, obeying a
three-term recurrence relation
\begin{equation}\label{eq:pnl_recurrence}
  \poly_{n+1,l}(s) = s\:\poly_{nl}(s) - \beta_{nl} \: \poly_{n-1,l}(s).
\end{equation}
In this way we only have to consider the single sequence of recurrence
coefficients $\beta_{nl}$. According to this normalisation the
$\rawpoly_{nl}(s)$ therefore obey the recurrence
\begin{equation}\label{eq:rawpnl_recurrence}
  \rawpoly_{n+1,l}(s) = -\imagi s\:\rawpoly_{nl}(s) + \beta_{nl} \: \rawpoly_{n-1,l}(s).
\end{equation}
Replacing $s$ with $\dop$ and applying to $\rho_{0l}$ on the right
then leads to the defining recurrence for the density basis elements
\eqref{eq:rhonl_recurrence}. Alternatively we can express the density
and potential directly in terms of $\poly_{nl}(s)$,
\begin{align}
  \Phi_{nlm} &= \imagi^{-n} p_{nl}\!\left(\imagi(r \partial_r + 1/2)\right) \Phi_{0lm}, \\ \nonumber
  \rho_{nlm} &= \imagi^{-n} p_{nl}\!\left(\imagi(r \partial_r + 5/2)\right) \rho_{0lm}.
\end{align}

\subsubsection{Disc case}\label{sec:disc_polynomials}

As expected, similar results apply in the case of \refedit{thin
  discs}. Take $\{\sigma_{nm}\}$ to be a set of
\refedit{infinitesimally thin} surface densities that are tridiagonal
with respect to the operator $\dopp$ and orthogonal with respect to
$\langle\cdot,\cdot\rangle$. We have index-raising polynomials
$\rawpoly_{nm}(s)$, defined by
\begin{equation}
  \rawpoly_{nm}(s) = \frac{\langle \Sigma_{s m}, \sigma_{nm} \rangle}
          {\langle \Sigma_{s m}, \sigma_{0m}\rangle}.
\end{equation}
This gives rise to a representation of the basis functions via
repeated application of the operator $\dopp$,
\begin{align}
  \sigma_{nm} &= \rawpoly_{nm}(\dopp) \sigma_{0m}, \\ \nonumber
  \psi_{nm} &= \rawpoly_{nm}(\dopp - \imagi) \psi_{0m}.
\end{align}
The orthogonality relation can be written
\begin{align}
  \langle \sigma_{nm}, \sigma_{n^\prime m^\prime} \rangle
  &= \delta_{m m^\prime} \int_{-\infty}^\infty ds \: \Omega_m(s) \:
    \rawpoly_{nm}(s) \: \overline{\rawpoly_{n^\prime m}(s)} \\ \nonumber
  &\propto \delta_{n n^\prime} \delta_{m m^\prime}
\end{align}
where
\begin{equation}
\label{eq:omega_m}
\Omega_m(s) = \frac{\left| \MellinTransform{\sigma_{0m}(R)}{R}{3/2 + \imagi s}
  \right|^2}{4 \pi K_m(\imagi s)}.
\end{equation}
The details of this derivation are in
App.~\ref{sec:disc_orth_proof}. As before we can instead use
real-valued polynomials $\poly_{nm}(s)$, also orthogonal with respect
to $\Omega_m(s)$. The potential and surface density in terms of
$\poly_{nm}(s)$ are
\begin{align}
  \psi_{nm} &= \imagi^{-n} p_{nm}\!\left(\imagi(R \partial_R + 1/2)\right) \psi_{0m}, \\ \nonumber
  \sigma_{nm} &= \imagi^{-n} p_{nm}\!\left(\imagi(R \partial_R + 3/2)\right) \sigma_{0m}.
\end{align}

In general it is difficult to find the $z$-dependence of the potential
for thin discs analytically, although in exceptional cases there may
be a simple solution (e.g.~the Kuzmin-Toomre discs). In any case
because $p_{nm}(\dopp)$ acts by differentiation with respect to $R$
alone, this guarantees that if the $z$-dependence of the zeroth-order
potential is known, then the correct the $z$-dependence is preserved
in all higher-order potential basis elements. This will have important
implications when considering the extension of our results to the
\citet{Ro96} method for thickened-disc basis sets, however we do not
pursue this in the present work.

\subsection{Completeness}\label{sec:completeness}

We make some informal comments about the completeness of a general
basis set $\{\rho_{nlm}\}$, derived from a zeroth-order $\rho_{0l}(r)$
as described above. The completeness of the angular part of each basis
(the spherical harmonics) is taken as given.

The question then of whether a set
$\{\rho_{0l}, \dop \rho_{0l}, \dop^2 \rho_{0l}, \ldots \}$ forms a
complete basis for (the $l$th multipole of) the space of mass
densities is the same as asking whether $\rho_{0l}$ is a \emph{cyclic
  vector} for the operator $\dop$. This is related to the completeness
of the associated orthogonal polynomials $p_{nl}(s)$, as powers of
$\dop$ correspond to powers of $s$; so we require that the monomials
$s^n$ (weighted by $\omega_l(s)$) form a complete basis for functions
on the interval $(-\infty,\infty)$. This is achieved if $\omega_l(s)$
is nonzero everywhere. By the definition of $\omega_l(s)$, this then
requires the Mellin transform of $\rho_{0l}$ to be nonzero everywhere,
which in turn requires that $\dop^n \rho_{0l}$ be non-vanishing
everywhere for all $n$ \citep{Marn2006}.

Therefore, to be a valid zeroth-order density, $\rho_{0l}$ must fulfil
the following:
\begin{align}\label{eq:valid_rho0l}
  \lVert \dop^n\rho_{0l} \rVert^2 &< \infty, \\ \nonumber
  \dop^n\rho_{0l}(r) &= 0 \: \text{at only isolated} \: r.
\end{align}
These conditions are required to hold for all $n \in \mathbb{N}$;
restricting to $n=0$ gives the conditions \eqref{eq:valid_rho} on
\emph{representable} mass densities. While these conditions are fairly
restrictive, in general any reasonable `analytical' potential-density
pairs will satisfy them; in particular those described in the
following section whose corresponding basis sets or index-raising
polynomials have closed-form expressions.

\section{Application to known basis sets}
\label{sec:existing}

In Sec.~\ref{sec:theory} we developed a theoretical justification for
the simple algorithm described in Sec.~\ref{sec:new}. We now provide
further motivation by applying the formalism to some concrete examples
of basis sets from the literature. Remarkably, all known analytical
spherical (resp.~\refedit{thin disc}) basis sets \refedit{of infinite
  extent} have a representation in terms of $\dop$ (resp.~$\dopp$). In
fact, it is extremely theoretically suggestive that \refedit{these}
previously-described \emph{analytical} basis sets have index-raising
polynomials that can be written in terms of known \emph{classical}
orthogonal polynomials. The expressions we derive below for the
various basis sets' index-raising polynomials may appear complicated;
however the presence of a classical polynomial indicates simply that
in each case the recurrence coefficient $\beta_{nl}$
\eqref{eq:rhonl_recurrence} can be written as a rational combination
of the given basis set's fixed shape parameters.

\subsection{Spherical case}
\label{sec:spherical}

\subsubsection{Clutton-Brock's Plummer basis set}\label{sec:cb73}

The simplest possible useful basis set in spherical polar coordinates
is that of \citet{CB73}, which uses the \citet{Plu1911} model as its
zeroth-order. By making an appropriate variable substitution,
Clutton-Brock transformed the Poisson equation for the radial
components \eqref{eq:spherical_basis_laplacian} into the defining
second-order differential equation for the Gegenbauer polynomials
\dlmf{18.8}. Each radial density and potential component is
proportional to just one polynomial,
\begin{align}
  \Phi_{nl}^\text{CB73}(r) &= \frac{-r^l}{(1+r^2)^{l+1/2}} C_n^{(l+1)}\!\!\left(\frac{r^2-1}{r^2+1}\right), \\ \nonumber
  \rho_{nl}^\text{CB73}(r) &= \frac{-(2n+2l+3)\: (2n+2l+1)\: \Phi_{nl}^\text{CB73}(r)}{4\pi (1+r^2)^2},
\end{align}
and the normalisation constant is\footnote{This corrects a typo in \citet{CB73}.}
\begin{equation}
  \int_0^\infty \!\! r^2 dr \: \Phi_{nl}^\text{CB73}(r) \rho_{nl}^\text{CB73}(r) =%
  -\frac{(2n\!+\!2l\!+\!3)(2n\!+\!2l\!+\!1)(n\!+\!2l\!+\!1)!}{2^{4l+6}(n+l+1)n!(l!)^2}.
\end{equation}
This basis set is in fact a special case of the family described in
Sec.~\ref{sec:lse_sets}, but as it is the simplest (and earliest) of
all the spherical basis sets we present it in some depth as a didactic
example.

Plugging $\rho_{0l}^{\text{CB73}}$ into the definition of the weight
function \eqref{eq:omega} we find that
\begin{align}
  \omega_l^\text{CB73}(s) &= 
    \frac{\left|%
  \Gamma\!\left(1/4 + l/2 + \imagi s/2\right)
  \Gamma\!\left(5/4 + l/2 + \imagi s/2\right) \right|^2}%
                            {8 \pi^2 \Gamma(l+1/2)^2} \\ \nonumber
  &= \frac{w\!\left(s/2;1/4 + l/2,5/4 + l/2\right)}{8 \pi^2 \Gamma(l+1/2)^2},
\end{align}
where $w(x;a,b)$ is the weight function for the \emph{continuous Hahn
  polynomials} $p_n(x;a,b)$ (App.~\ref{sec:continuous_hahn}). It can
be verified that each basis element $\rho_{nl}^\text{CB73}$ and
$\Phi_{nl}^\text{CB73}$ indeed Fourier-Mellin-transforms into a single
continuous Hahn polynomial. Specifically, we find that
\begin{align}
\label{eq:CB73_Ps}
  \rawpoly_{nl}^\text{CB73}(s) &= A_{nl} \: \imagi^n \: p_n\!\left(\frac{s}{2}; \frac{1}{4} + \frac{l}{2}, \frac{5}{4} + \frac{l}{2}\right), \\ \nonumber
  A_{nl} &= \frac{\sqrt{\pi} \: \Gamma(l+1/2) \: (n+2l+1)!}{2^{2l} \: (2n+2l+1) \: l! \: \Gamma(n+l+1/2)^2}.
\end{align}
Looking at the definition of a continuous Hahn polynomial
\eqref{eq:continuous_hahn_def}, we find a hypergeometric function that
terminates after $n$ terms, but where the argument $s$ appears as a
`parameter'. Given how this relates to the definition of the density
elements, this means that
\begin{align}
\label{eq:CB73_Ps_explicit}\nonumber
  \rho_{nl}^\text{CB73}  &\!=\! B_{nl} \: \hypergeom{3}{2}{-n, n+2l+2, l/2 + 1/4 + \imagi \dop/2}{l+1/2, l+3/2}{1} \rho_{0l}^\text{CB73}, \\
  B_{nl} &= (-1)^n\!\properbinom{n\!+\!2l\!+\!1}{n},
\end{align}
where the operator $\dop$ alarmingly also appears as a `parameter';
each term in the sum is proportional to a Pochhammer symbol whose
argument involves $\dop$. However these are unproblematic to evaluate,
as they expand to
\refedit{%
\begin{multline}
  \left(l/2 + 1/4 + \imagi \dop/2\right)_j
  = \left(l/2 - 1 - r \partial_r/2 + j-1\right)\times\\
  \left(l/2 - 1 - r \partial_r/2 + j-2\right)\times\ldots\times
  \left(l/2 - 1 - r \partial_r/2\right),
\end{multline}}
and each occurrence of $r \partial_r$ then operates to the right on
$\rho_{0l}^\text{CB73}(r)$ in the expected fashion. The index-raising
polynomials (of argument $\dop$ or $\dopp$) in the remainder of this
section are evaluated in a similar way.

\subsubsection{The double-power law basis sets}
\label{sec:lse_sets}

Practically all known double-power law basis sets in spherical polar
coordinates are contained within one super-family described in
\citet{LSE} (containing within it the basis sets of
\citet{CB73,HO92,Zh96,Ra09,LSEE}). There are two free parameters
($\alpha$ and $\nu$) controlling both the asymptotic power-law slope
and turnover. We refer to the expressions given in \citet{LSE} for the
potential, density and normalisation constants (Eqs~(30)--(33) of that
work), and label them with the superscript $\mathrm{LSE}$. The
zeroth-order has $\rho_{0l}^{\text{LSE}} \sim r^{-2+1/\alpha+l}$ as
$r \to 0$, and $\rho_{0l}^{\text{LSE}} \sim r^{-3-\nu/\alpha-l}$ as
$r \to \infty$. Inserting $\rho_{0l}^{\text{LSE}}$ into the definition
of the weight function \eqref{eq:omega} and writing
$\mu = \alpha(1+2l)$, we find that
\begin{equation}
  \omega_l^\text{LSE}(s) =
  \left(\frac{\mu}{4\pi\Gamma(\mu+\nu)}\right)^2 \:
  w\!\left(\alpha s; \frac{\mu}{2}, \frac{\mu}{2} + \nu\right),
\end{equation}
which is again proportional to a continuous Hahn weight function
(App.~\ref{sec:continuous_hahn}). Explicitly for the index-raising
polynomials we have
\begin{align}
\label{eq:lse_Ps}
\rawpoly_{nl}^\text{LSE}(s) &= \frac{K_{nl}^\text{LSE}}{K_{0l}^\text{LSE}}
\frac{\imagi^n n!}{(\mu)_n (\mu+\nu)_n}
p_n\!\left(\alpha s; \frac{\mu}{2}, \frac{\mu}{2} + \nu\right) \\ \nonumber
&= \frac{(-1)^n K_{nl}^\text{LSE}}{K_{0l}^\text{LSE}}
\hypergeom{3}{2}{-n, n+2\mu+2\nu-1, \mu/2
  + \imagi \alpha s}{\mu, \mu+\nu}{1}.
\end{align}

\subsubsection{The cuspy-exponential basis sets}
\label{sec:cuspy_exponential}

These basis sets were not mentioned in \citet{LSE} and are therefore
newly presented in the literature\footnote{See
  \citet[Ch.~6]{lilley_thesis} for a detailed derivation.}; but they
are a straightforward derivation from the double power-law result,
obtained by letting the parameter $\nu$ and the scalelength
simultaneously tend to infinity. The result is a family of basis sets
with both an exponential fall-off and a central cusp in density, both
controlled by the parameter $\alpha$ -- hence the nickname \emph{cuspy
  exponential}. The lowest order density function is
$\rho_{00} \propto r^{-2+1/\alpha}\expe^{-r^{1/\alpha}}$. Important
cases are $\alpha = 1/2$ which gives a Gaussian, and $\alpha = 1$
which is a density familiar to chemists as the Slater-type orbital. We
use the superscript $\text{CE}$ for these basis functions. The density
and potential are \refedit{(with $\mu = \alpha(1+2l)$)}
\begin{align} \nonumber
  \rho_{nl}^\text{CE}(r) &= 2 \: (-1)^n r^{l-2+1/\alpha}\expe^{-r^{1/\alpha}} \!\!
  \left[ L_n^{(\mu)}\left(2r^{1/\alpha}\right)
    + L_{n-1}^{(\mu)}\left(2r^{1/\alpha}\right)\right], \\ \nonumber
  \Phi_{0l}^\text{CE}(r) &= \frac{\mu \: \upgamma\!
    \left(\mu,r^{1/\alpha}\right)}{r^{l+1}}, \\ 
  \Phi_{nl}^\text{CE} &- \Phi_{n+1,l}^\text{CE} =\frac{2 n! (-1)^n}{(\mu+1)_n}
  r^l \expe^{-r^{1/\alpha}} L_n^{(\mu)}\left(2r^{1/\alpha}\right),
\end{align}
\refedit{where $\upgamma(\mu,z)$ is the (lower) incomplete Gamma function and
$L_n^{(\mu)}(z)$ is a Laguerre polynomial}. The relevant constants are
\begin{align}
  N_{nl}^\text{CE} &= \frac{\alpha \Gamma(\mu+1)}{2^{\mu-1}}, \\ \nonumber
  K_{nl}^\text{CE} &= \frac{-n! \Gamma(\mu+1)}{8\pi\alpha^2 \Gamma(n+\mu)}.
\end{align}
We can apply the limiting procedure directly to
$\rawpoly_{nl}^\text{LSE}(s)$, and the calculation is simpler than for
the basis functions themselves. The operator $\dop$ does not depend on
the scalelength, and hence is unaffected by the limiting procedure. So
we need only consider the limit in $\nu$.  The result is proportional
to a \emph{Meixner-Pollaczek polynomial} $P_n^{(\mu/2)}(z;\phi)$
(App.~\ref{sec:meixner_pollaczek}),
\begin{align}
  \rawpoly_{nl}^\text{CE}(s) &= \frac{K_{nl}^\text{CE}}{K_{0l}^\text{CE}}
  \frac{\imagi^n n!}{(\mu)_n} P_n^{(\mu/2)}\left(\alpha s; \frac{\pi}{2}\right).
\end{align}

\subsection{\refedit{Thin disc} case}
\label{sec:disc}

\subsubsection{Clutton-Brock's Kuzmin-Toomre basis set}\label{sec:cb72}

The Kuzmin-Toomre model \citep{Kuz56,Too63} is the simplest power law
for the infinitesimally thin disc. This model provides the
zeroth-order for a basis set introduced by \citet{CB72}. This basis
set turns out to be a special case of Qian's family
(Sec.~\ref{sec:qian-k}), but here at least we can write down simple
expressions in terms of a single Gegenbauer polynomial
$C_n^{(\alpha)}(x)$ \citep{Ao78}, so it is worth recording the results
separately. The density and potentials in the plane are
\begin{align}
  \psi_{nm}^\text{CB72}(R,\varphi) &= 
  \frac{-\expe^{\imagi m \varphi} \: R^m}{\left(1 + R^2\right)^{m+1/2}} C_n^{(m+1/2)}\!
  \left(\frac{R^2-1}{R^2+1}\right), \\ \nonumber
  \sigma_{nm}^\text{CB72}(R,\varphi) &= -\frac{n+m+1/2}{2\pi} \:
  \frac{\psi_{nm}^\text{CB72}(R,\varphi)}{1 + R^2},
\end{align}
and the normalisation constant in the orthogonality relation is
\begin{equation}
N_{nm}^\text{CB72}  = \intdd \: \sigma_{nm}^\text{CB72} \: \psi_{nm}^\text{CB72}
= \frac{-\pi \: \Gamma(n + 2m + 1)}{2^{4m+2} \: n! \: \Gamma(m + 1/2)^2}.
\end{equation}
The corresponding Fourier-Mellin weight function \eqref{eq:omega_m} is
then
\begin{equation}
  \Omega_m^\text{CB72}(s) = \frac{\left| \Gamma(1/2 + m + \imagi s) \right|^2}
        {2^{2m+5} \pi^2 \Gamma(m + 1/2)^2},
\end{equation}
which is proportional to the weight function for a Meixner-Pollaczek
polynomial (App.~\ref{sec:meixner_pollaczek}) with parameters
$\lambda = m + 1/2$ and $\phi = \pi/2$. So the index-raising
polynomials have a simple expression in terms of the Meixner-Pollaczek
polynomials,
\begin{align}
  \rawpoly_{nm}^\text{CB72}(s) &= \imagi^n \: P_n^{(m+1/2)}(s; \pi/2). %
\end{align}

\subsubsection{Qian's $k$-basis sets}\label{sec:qian-k}

The family of basis sets introduced by \citet{Qi93} is a
generalisation of \citet{CB72}, allowing for an arbitrary
\textit{generalised} Kuzmin-Toomre model to be the zeroth order. That
is, the zeroth-order density functions are (using the superscript
$\text{Q}$)
\begin{align}\label{eq:k_basis_sigma_psi}
  \sigma_{0m}^\text{Q}(R,\varphi) &= \sqrt{\pi}\frac{\Gamma(m+k+3/2)}{\Gamma(m + k + 1)} \frac{R^m \: \expe^{\imagi m \varphi}}{\left(1 + R^2\right)^{m+k+3/2}}. \\ \nonumber
  \psi_{0m}^\text{Q}(R,\varphi) &= \pi \: \Beta\!\left(m+\frac{1}{2},\frac{1}{2}\right) \: \frac{R^m \: \expe^{\imagi m \varphi}}{(1 + R^2)^{m+1/2}} \\ \nonumber
                                  &\qquad \times \hypergeom{2}{1}{-k,m+1/2}{m+1}{\frac{R^2}{1 + R^2}}.
\end{align}
\refedit{Here $\Beta(x,y) = \Gamma(x)\Gamma(y)/\Gamma(x+y)$ is the standard
Beta function, and the} prefactors have been chosen so that all derived
expressions are compatible with those in \citet{Qi93}. The
higher-order potential and density functions that Qian provides are
given in terms of very complicated recursion relations, that are only
valid when $k$ is an integer. %
However there is no such limitation in our representation. The weight
function is proportional to that for a continuous Hahn polynomial
$p_n(s;a,b)$ (App. \ref{sec:continuous_hahn}), and so the
index-raising polynomial is
\refedit{
\begin{equation}
  \label{eq:k_basis_Pnm}
  \rawpoly_{nm}^\text{Q}(s) =
  \frac{\imagi^{n}}{(m+k+1)_n} \:
  p_n\!\left(\frac{s}{2}; \frac{m}{2}+\frac{1}{4},
  \frac{m}{2}+k+\frac{3}{4}\right).
\end{equation}}
We therefore have closed-form expressions for $\sigma_{nm}^\text{Q}$
and $\psi_{nm}^\text{Q}$, that are valid for all real values of $k$,
as long as the zeroth-order model has finite total self-energy. The
original \citet{CB72} basis set (Sec.~\ref{sec:cb72}) is recovered
when $k=0$. The normalisation constant for the orthogonality relation
can be derived from that of the continuous Hahn polynomials, and is
\begin{align}
  N_{nm}^\text{Q} &= \int_0^\infty R \: dR \: \sigma_{nm}^\text{Q} \: \psi_{nm}^\text{Q} \\ \nonumber
  &=    \frac{\pi ^2 \Gamma \left(m+n+\frac{1}{2}\right) \Gamma \left(2
          k+m+n+\frac{3}{2}\right)}{2 n! (2 k+2 m+2 n+1) \Gamma (2 k+2 m+n+1)}.
\end{align}

\subsubsection{Qian's Gaussian basis set}

A Gaussian density profile is another plausible model for the density
of a galactic disc, and such a basis set was also studied by
\citet{Qi93}. Just as we derived the cuspy-exponential basis sets of
Sec.~\ref{sec:cuspy_exponential} from the double-power law result by
taking the infinite limit of the shape parameter $\nu$, it turns out
that Qian's basis set for the Gaussian disc can be derived by taking
the limit $k \to \infty$ in the corresponding expressions
\eqref{eq:k_basis_sigma_psi} for the generalised Kuzmin-Toomre basis
set of Sec.~\ref{sec:qian-k}. The zeroth-order density and potential
are (using the superscript $\text{G}$)
\begin{align}\label{eq:G_basis_sigma_psi}
  \sigma_{0m}^\text{G}(R,\varphi) &= \lim_{k \to \infty}\left\{k^{\frac{m}{2}} \sigma_{0m}^\text{Q}\!\left(\frac{R}{\sqrt{k}},\varphi\right)\right\} \\ \nonumber
                                  &= \sqrt{\pi} \: R^m \: \expe^{-R^2} \: \expe^{\imagi m \varphi}, \\ \nonumber
  \psi_{0m}^\text{G}(R,\varphi) &= \lim_{k \to \infty}\left\{k^{\frac{m}{2}} \psi_{0m}^\text{Q}\!\left(\frac{R}{\sqrt{k}},\varphi\right)\right\} \\ \nonumber
  &= \pi \: \Beta\!\left(m+\frac{1}{2},\frac{1}{2}\right) \: R^m \: \expe^{\imagi m \varphi} \hypergeom{1}{1}{m+1/2}{m+1}{-R^2}.
\end{align}
The function denoted ${}_1F_1$ is a confluent hypergeometric (Kummer)
function, that reduces to combinations of modified Bessel functions
for any given $m$. At zeroth-order we have the well-known result that
the potential of a plain Gaussian disc involves a single modified
Bessel function,
$\psi_{00}^\text{G}(R) = \pi^2 I_0\!\left(R^2/2\right)\expe^{-R^2/2}$.

Again Qian gives the higher-order potential and densities only as
complicated recursion relations. However, explicit expressions follow
upon taking the limit $k \to \infty$ in \eqref{eq:k_basis_Pnm}. We
find that
\begin{align}\label{eq:G_basis_Pnm}
  \rawpoly_{nm}^\text{G}(s) &= \lim_{k \to \infty}\left\{\rawpoly_{nm}^\text{Q}(s)\right\} = \imagi^{-n} \: P_n^{(m/2 + 1/4)}\!\left(\frac{s}{2}; \frac{\pi}{2}\right), \\ \nonumber
  N_{nm}^\text{G} &= \lim_{k \to \infty}\left\{k^{m+\frac{1}{2}} N_{nm}^\text{Q}\right\} = \frac{\pi^2 \Gamma(n+m+1/2)}{2^{m+3/2} n!},
\end{align}
where $P_n^{(m/2+1/4)}(s/2;\pi/2)$ is a Meixner-Pollaczek polynomial
(Sec.~\ref{sec:meixner_pollaczek}). Then
\eqref{eq:G_basis_sigma_psi} and \eqref{eq:G_basis_Pnm} can be
combined to find
\begin{align}
  \sigma_{nm}^\text{G}(R,\varphi) &= \lim_{k \to \infty}\left\{k^{\frac{m}{2}} \sigma_{nm}^\text{Q}\!\left(\frac{R}{\sqrt{k}},\varphi\right)\right\} \\ \nonumber
                                  &= \lim_{k \to \infty}\left\{k^{\frac{m}{2}} \rawpoly_{nm}^\text{Q}(\dopp) \sigma_{0m}^\text{Q}\!\left(\frac{R}{\sqrt{k}},\varphi\right)\right\} \\ \nonumber
                                  &= \rawpoly_{nm}^\text{G}(\dopp) \sigma_{0m}^\text{G}\!\left(R,\varphi\right),
\end{align}
which works because the factor of $1/\sqrt{k}$ cancels out in $\dopp$;
there is a similar expression for the potential functions.

\subsubsection{Exponential disc}\label{sec:expdisc}

Interestingly, there is another \refedit{thin disc} model which has
classical index-raising polynomials: we briefly sketch the derivation
for an exponential disc.

\begin{figure}
  \includegraphics[width=0.5\textwidth]{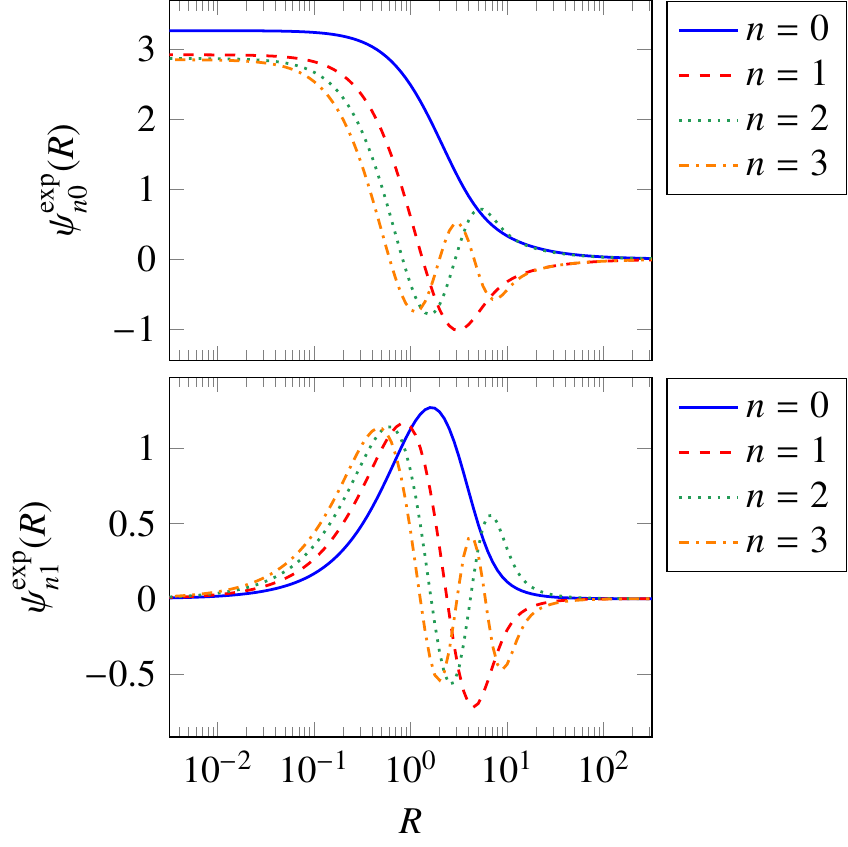}
  \caption{Radial parts of the exponential disc basis
    $\psi^\text{exp}_{nm}(r)$ for $n = 0,1,2,3$ and $m=0$ (top) and
    $m=1$ (bottom). The potentials have been unit-normalised.}
  \label{fig:expdisc_pot}
\end{figure}

We require all density components to fall off exponentially like
$\expe^{-R}$ but also to behave like an interior multipole as
$R\to 0$, so as a zeroth-order ansatz for the density we take simply
\begin{equation}\label{eq:exp_disc_sigma}
  \sigma_{0m}^\text{exp}(R) = R^m \expe^{-R}.
\end{equation}
This gives a weight function (via \eqref{eq:omega_m}) proportional to
that for a continuous Hahn polynomial\footnote{Unfortunately
  generalising the exponent to $\expe^{-R^{1/\alpha}}$ gives no
  similarly simple result.}. Thus the index-raising polynomials can be
written down explicitly as
\begin{equation}
  P_{nm}^\text{exp}(s) = \imagi^{-n} \: p_n\!\left(s/2; m/2+1/4, m/2+5/4\right),
\end{equation}
along with closed-form expressions for the recurrence coefficient and
normalisation constant. The remaining complication is the zeroth-order
potential. The $m=0$ case is awkward but classical \citep[Ch. 2]{BT} \refedit{and uses modified Bessel functions},
\begin{equation}\label{eq:exp_disc_psi0}
\psi_{00}^\text{exp}(R) = -\pi R\!\left[ I_0\!\left(\frac{R}{2}\right)\!K_1\!\left(\frac{R}{2}\right) - I_1\!\left(\frac{R}{2}\right)\!K_0\!\left(\frac{R}{2}\right) \right].
\end{equation}
Deriving expressions when $m\!>\!0$ is trickier -- we give the details
in App.~\ref{app:exponential_disc} -- but it can be accomplished with
the following differential-recurrence relation:
\begin{equation}\label{eq:exp_disc_psi}
\psi_{0,m+1}^\text{exp}(R) = \frac{\left( R \partial_R - m - 1 \right)\left( \partial_R - m/R \right)}{2m+3} \psi_{0m}^\text{exp}(R).
\end{equation}
Some examples of the potential basis elements
$\psi_{nm}^\text{exp}(R)$ are plotted in Fig.~\ref{fig:expdisc_pot}.

\section{Numerical implementation}\label{sec:numerical}

At the end of Sec.~\ref{sec:new} we mentioned the main obstacles to
the effective implementation of the new algorithm -- primarily the
numerical stability when computing the coefficients $\beta_{nl}$, but
also the need to compute repeated radial derivatives of the
zeroth-order elements.

For the recurrence coefficients $\beta_{nl}$ the difficulty is that
naively computing the integrals \eqref{eq:betanl_def} becomes
computationally expensive very quickly with increasing order $n$ (and
to some extent also with $l$). Therefore it is essential to pick a
numerical integration method that is fast without sacrificing
accuracy. Unfortunately due to the total freedom in choice of
zeroth-order $\rho_{0l}$, it is difficult to find a quadrature scheme
for the integrals \eqref{eq:betanl_def} that is optimal in general.

Fortunately, due to the link to the polynomials $\poly_{nl}(s)$
developed in Sec.~\ref{sec:theory}, we can take advantage of the
extensive literature on the construction of general orthogonal
polynomials. Following \citet{Gautschi1985} we have two options:
either the \emph{discretized Stieltjes procedure} or the
\emph{modified Chebyshev algorithm}. As it happens, computing the
recurrence coefficients naively as in \eqref{eq:betanl_def} is
directly analogous to using the discretized Stieltjes procedure,
except that now we perform the integrals in Fourier-Mellin space.
This turns out to be the better option numerically, as the modified
Chebyshev algorithm runs into floating point issues sooner due to
catastrophic cancellation of terms. However, for completeness we
describe both algorithms (Sections \ref{sec:discretized_stieltjes} and
\ref{sec:modified_chebyshev}). We also discuss computer-assisted
techniques for performing the repeated differentiations
(Sec.~\ref{sec:repeated_diff}).

All these methods are illustrated throughout for a basis set
constructed to have the \emph{isochrone model} \citep{Henon59} as its
zeroth-order, and we follow up the numerical discussion with a
demonstration of the validity of the isochrone-adapted basis set
(Sec.~\ref{sec:application}); however the underlying methods we
describe are applicable to any suitable zeroth-order model. The
potential, density and polynomial weight function for the isochrone
model are as follows:
\begin{align}\label{eq:isochrone}
\Phi_{0l}^\text{iso}(r) &= \frac{-r^l}{\left(1 + \sqrt{1 + r^2}\right)^{2l+1}}, \\ \nonumber
\rho_{0l}^\text{iso}(r) &= \frac{(2l+1)r^l\left[(2l+3)(1 + \sqrt{1+r^2}) + (2l+2)r^2\right]}{(1 + \sqrt{1+r^2})^{2l+3}(1+r^2)^{3/2}}, \\ \nonumber
\omega_l^\text{iso}(s) &= \frac{(2l+1)^2}{2^{2l+6} \pi^2} \left| \frac{\Gamma\!\left(l + 3/2 + \imagi s\right) \Gamma\!\left(l/2 + 1/4 + \imagi s/2\right)}{\Gamma\!\left(3l/2 + 7/4 + \imagi s/2\right)}\right|^2.
\end{align}
The precise $l$-dependence of these expressions is of course arbitrary
to some extent, but we have made a suitable `natural' choice.

\subsection{Discretized Stieltjes procedure}\label{sec:discretized_stieltjes}

The sequence of recurrence coefficients $\beta_{nl}$ that we need to
compute can be expressed as the ratio of two integrals,
\refedit{%
\begin{equation}\label{eq:beta_DS}
  \beta_{nl} = \frac{I_{nl}}{I_{n-1,l}}, \:\: \mathrm{where} \:\: I_{nl} = \langle \rho_{nl}, \rho_{nl} \rangle = \lVert \rho_{nl} \rVert^2,
\end{equation}}
and so for each higher $n$ we need one additional evaluation of
$I_{nl}$.  Evaluations of $\beta_{nl}$ alternate with applications of
the recurrence relation \eqref{eq:pnl_recurrence} to find the next
basis element $\rho_{nl}$. Once sufficient $\beta_{nl}$ have been
found, the potential or density functions $\rho_{nlm}$ and
$\Phi_{nlm}$ can be evaluated via their own recurrences as described
in Sec.~\ref{sec:new}.

The difficulty then is in finding an appropriate strategy to compute
the integrals $I_{nl}$. We opt to evaluate them in
Fourier-Mellin-space, using the polynomials $p_{nl}(s)$ directly, and
making use of the fact that the integral can be written
\begin{equation}\label{eq:Inl_poly_integral}
I_{nl} = \int_{-\infty}^\infty \omega_l(s)  \left[p_{nl}(s)\right]^2  ds.
\end{equation}
Therefore the first step is to determine the weight function
$\omega_l(s)$. This can be found in terms of the (Fourier-)Mellin
transform of either the zeroth-order potential or the density,
\begin{align}\label{eq:omega_l_expressions}
  \omega_l(s) &= \frac{2 K_{0l}^2}{(l + 1/2)^2 + s^2} \left|\MellinTransform{\rho_{0l}(r)}{r}{5/2 + \imagi s} \right|^2 \\ \nonumber
              &= \frac{(l + 1/2)^2 + s^2}{8\pi^2} \left|\mathcal{M}_r\left\{\Phi_{0l}(r)\right\}(1/2 + \imagi s)\right|^2.
\end{align}
The Mellin transform is perhaps one of the less familiar integral
transforms, but in practice a wide variety of Mellin transforms can be
found in closed-form (helped especially by computer algebra systems),
in part because with a logarithmic change of variable it can be
written as a Fourier transform. \refedit{All the polynomial weight
  functions considered in this paper can be found symbolically using
  \textsc{Mathematica}\footnote{\refedit{Some \textsc{Mathematica}
      code demonstrating this is included in the repository at
      \url{https://github.com/ejlilley/basis}.}}. Numerical evaluation
  of the Mellin transform is also an option -- by transformation to
  the Fourier transform and approximation using Fast Fourier Transform
  methods -- however we do not pursue this further in the present
  work.}

Now we consider the asymptotic behaviour of the weight function
$\omega_l(s)$ as $s \to \pm\infty$. The smoothness requirement
\eqref{eq:valid_rho} on $\rho_{0l}$ forces $\omega_l(s)$ to decay
faster than any power of $s$, i.e.~at least exponentially.
\refedit{We expect that
\begin{equation}
  \omega_l(s) \sim |s|^b \: \expe^{-a|s|} \:\: \mathrm{as} \:\: s \to \pm\infty,
\end{equation}
so we need to determine the decay constant $a$.} In the case of our
isochrone basis set this asymptotic behaviour is derived from the
behaviour of the complex gamma function at infinity \dlmf{5.11.9},
giving
$\omega_l(s) \sim \left|s\right|^{-1} \expe^{-\pi\left|s\right|}$, or $a = \pi$.
When $\omega_l(s)$ can be written down, it is usually simple to read
off the decay constant $a$; for example, the double power-law basis
sets (Sec.~\ref{sec:lse_sets}) have $a = \alpha$.

The at-least-exponential decay of the weight function
suggests that the appropriate discretisation scheme for
\eqref{eq:Inl_poly_integral} is \emph{Gauss-Laguerre quadrature}. To
implement this for the isochrone case, rewrite
\eqref{eq:Inl_poly_integral} to pull out a factor of $\expe^{-\pi s}$,
and use the symmetry of the integrand to change the domain of
integration to $(0,\infty)$ (defining $x = \pi s$),
\begin{equation}
  I_{nl} = \frac{2}{\pi} \int_0^\infty \expe^{-x} \: \underbrace{\expe^{x} \: \omega_l(x/\pi) \: \left[p_{nl}(x/\pi)\right]^2}_{\sim x^{2n-1}\text{ as }x\to \infty} \: dx
\end{equation}
We can then implement Gauss-Laguerre quadrature of order $\nu$, as a
weighted sum over evaluation points $x_{j\nu}$, which are the roots of
the $\nu$th Laguerre polynomial $L_\nu(x)$:
\begin{align}\label{eqn:gauss_laguerre}
  I_{nl} \approx \frac{2}{\pi} \sum_{j=1}^\nu w_{j\nu} \: \expe^{x_{j\nu}} \: \omega_l(x_{j\nu}/\pi) \: \left[p_{nl}(x_{j\nu}/\pi)\right]^2, \\ \nonumber
  w_{j\nu} = \frac{x_{j\nu}}{\left[(\nu+1) \: L_{\nu+1}\left(x_{j\nu}\right)\right]^2}, \\ \nonumber
  L_\nu\!\left(x_{j\nu}\right) = 0, \qquad j = 1\ldots \nu.
\end{align}
The quadrature rule of order $\nu$ integrates polynomials exactly up
to order $2\nu-1$, so to compute $I_{nl}$ with the isochrone weight
function we would expect to need at least $\nu \geq n$. (An acceptable
rule of thumb is that $I_{nl}$ requires $\nu = \text{max}(n+l,10)$.)
It may be necessary to compute the weights $w_{j\nu}$ and roots
$x_{j\nu}$ to a higher order of precision internally using
arbitrary-precision arithmetic, but this is not a bottleneck in
practice -- and typically Gauss-Laguerre quadrature is implemented as
a library function whose implementation details are hidden.  In this
way we can get e.g.~$50$ orders of $\beta_{nl}$ to floating-point
precision in under a tenth of a second using one core of a modern CPU.

\begin{figure}
  \includegraphics[width=0.5\textwidth]{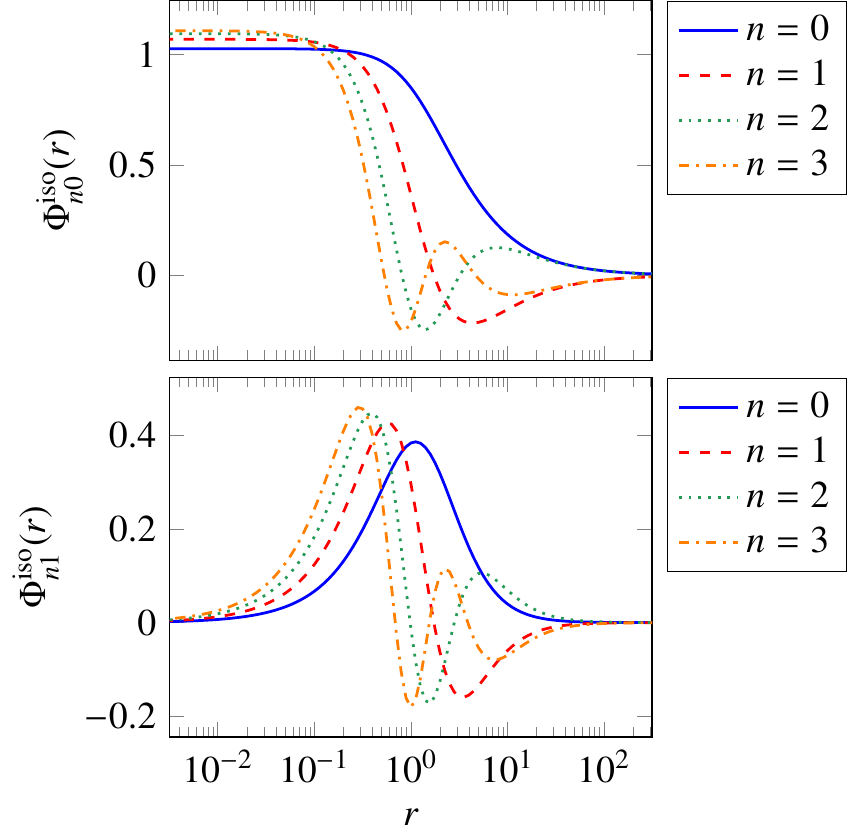}
  \caption{Radial parts of the isochrone potential basis
    $\Phi^\text{iso}_{nl}(r)$ for $n = 0,1,2,3$ and $l=0$ (top) and
    $l=1$ (bottom). The potentials have been unit-normalised.}
  \label{fig:isochrone_pot}
\end{figure}

The radial parts of some examples of potential elements in the
isochrone basis set are plotted in Fig.~\ref{fig:isochrone_pot}.

\subsection{Modified Chebyshev algorithm}\label{sec:modified_chebyshev}

This is an alternative method described in \citet{Gautschi1985}, which
we find to be less numerically stable in practice. However we describe
it here for completeness, as it may yet find some usefulness (e.g.~to
facilitate finding exact expressions for the recurrence coefficients
in certain cases).
Gautschi's modified Chebyshev algorithm prescribes the
\emph{modified moments}\footnote{This is in distinction to Chebyshev's
  original algorithm, which uses the raw moments
  $\mu_k = \int s^k \: \omega(s) \: ds$.}
\begin{equation}
  {\tilde\mu}_{kl} \equiv \int_{-\infty}^\infty \omega_l(s) \: {\tilde p}_{kl}(s).
\end{equation}
Here, ${\tilde p}_{kl}(s)$ are some auxiliary set of (monic)
polynomials, orthogonal with respect to a symmetric measure on the
interval $(-\infty,\infty)$, and obeying a three-term recurrence relation
\begin{align}\label{eq:ptkl_recurrence}
  {\tilde p}_{-1,l}(s) &= 0, \\ \nonumber
  {\tilde p}_{k+1,l}(s) &= s\:{\tilde p}_{kl}(s) - {\tilde\beta}_{kl}{\tilde p}_{k-1,l}(s).
\end{align}
By symmetry this means that ${\tilde\mu}_{kl}$ are nonzero only for
even $k$. In principle the choice of auxiliary polynomial is wide
open, but the obvious choice in our case (for ease and stability of
computation) is the monic Hermite polynomials $H\!e_k(s)$, for which
${\tilde \beta}_{kl} = k$. We can then proceed to find the \emph{mixed}
moments
\begin{equation}
  \sigma_{jkl} = \int_{-\infty}^\infty p_{jl}(s) \: {\tilde p}_{kl}(s) \: \omega_l(s) \: ds
\end{equation}
via a system of recurrence relations that produces the desired
recurrence coefficients $\beta_{nl}$ as a byproduct:
\begin{align}
  \sigma_{0kl} &= {\tilde \mu}_{kl}, \\ \nonumber
  \sigma_{jkl} &= \sigma_{j-1,k+1,l} - \beta_{j-1,l} \sigma_{j-2,k,l} + {\tilde \beta}_{kl} \sigma_{j-1,k-1,l}, \\ \nonumber
  \beta_{nl} &= \frac{\sigma_{nnl}}{\sigma_{n-1,n-1,l}}.
\end{align}
In practice (for our isochrone basis set) we find that $\sigma_{jkl}$
suffers from catastrophic cancellation beyond approximately
$j=20$. Alternatively if the modified moments ${\tilde \mu}_{kl}$ are
known in `closed-form' then this method is convenient for finding
`exact' recurrence coefficients. This turns out to be the case for
the isochrone basis set, for which see App.~\ref{sec:isochrone_exact}.

\subsection{Repeated differentiation}\label{sec:repeated_diff}

There are three classes of algorithm for computer-assisted
differentiation: \begin{inparaenum}
\item finite-differencing,
\item symbolic differentiation, and
\item automatic differentiation.
\end{inparaenum}
The first of these we can discount pretty much immediately as being
wildly numerically unstable and expensive compared to the other
two.

The second, symbolic differentiation via computer algebra, is
potentially competitive at low expansion orders, but it is hard to
predict the degree of blow-up in the number of algebraic terms. It
depends strongly on the precise form of the function that is being
differentiated. In practice we find that efficient application of
symbolic differentiation at high expansion orders requires alternating
between differentiation and algebraic simplification.

Of course, one may also attempt symbolic differentiation by hand,
attempting to find simplifications that reduce the tower of
applications of $\dop^n$ to a simpler form -- whether this is possible
also depends on the form of $\Phi_{0l}$ and $\rho_{0l}$. Many of the
basis sets considered in Sec.~\ref{sec:existing} have simple
closed-forms (at all orders) due to fortunate simplification in
repeated differentiation. For example, taking the double-power law
basis (Sec.~\ref{sec:lse_sets}) with parameters $\alpha = 1/2$,
$\nu = p - 3/2$ and $n=l=0$ (and labelling each density function with
$p$), we have%
\begin{equation}
\rho_{p+1} = p^{-1} \: (p - 5/4 - \imagi\dop/2) \rho_p.
\end{equation}
Using this identity in \eqref{eq:lse_Ps} then leads (after some
further simplification) to a known closed-form expression for
$\rho_{nl}^\text{LSE}$. However it is likely to be difficult to find
easy differentiation formulas in general. For our isochrone basis set,
the method we give in App.~\ref{sec:isochrone_exact} for computing the
modified moments can be adapted to find expressions for the
higher-order derivatives, but the result is complicated and of dubious
numerical stability.

The third method, automatic differentiation (AD) is what we find to be
most competitive in practice. This is a general term referring to a
class of algorithms implemented entirely at the software library
level, that provides an evaluation of the derivative at a single point
given only knowledge of the chain rule and the differentiation rules
for primitive arithmetic operations and standard mathematical library
functions. Essentially, the function to be differentiated is written
in ordinary code, and the AD algorithm \emph{automatically} deduces
the correct sequence of chain rule steps to carry out. For our
purposes we require higher-order derivatives; while applying an AD
algorithm to itself works in principle (and often works in practice)
it is very inefficient, as the AD logic itself must be
differentiated. It is better to use an AD implementation that natively
understands higher-order derivatives.

As we are coding in the \textsc{Julia} programming language, we use a
suitable library called \textsc{TaylorSeries.jl}
\citep{TaylorSeries.jl}. A special variable $t(N)$ is instantiated
that represents the first $N$ terms of an (abstract) Taylor
series. Given a point $r_0$, we can use $t + r_0$ as the argument of
any ordinary mathematical function\footnote{I.e.~a function that
  accepts and returns a floating-point value.}; the result is the
first $N$ coefficients of the Taylor series around $r_0$ that
approximates that function. For example, setting $N=3$ and $r_0 = 1.0$
and using the potential of the isochrone model \eqref{eq:isochrone} as
our function, the computer prints a data structure representing the
following truncated Taylor series,
\refedit{%
\begin{multline}\nonumber
  \Phi_{00}^\text{iso}(t(3) + 1.0) =  - 0.4142 + 0.1213 t - 0.0052 t^2 - 0.0225 t^3. %
\end{multline}}

When it comes to the actual implementation, we have two choices, which
we find to have similar efficiency in practice. The first option
begins with computing the vector of derivatives (at a point
$\vec{r}_0$) up to some maximum order $N$ all in one go,
\begin{equation}
  \vec{V}_l = (\Phi_{0lm}, \dop \Phi_{0lm}, \dop^2 \Phi_{0lm}, \ldots, \dop^N \Phi_{0lm}).
\end{equation}
In fact, because $\dop$ can be expressed as a single differentiation
with respect to a transformed variable (via $r \: d/dr = d/ds$, where
$s = \log r$), $\vec{V}_l$ can be obtained directly from a single
$N$-term Taylor series evaluation. Separately, we derive from
$\beta_{nl}$ the matrix elements $(\mathsf{A}_l)_{nj} = A_{njl}$ in
the expansion
\begin{equation}
  \Phi_{nlm} = \sum_{j=0}^n A_{njl} \: \dop^j \Phi_{0lm}.
\end{equation}
To evaluate a vector of potential functions at a single point,
\begin{equation}
\vec{\Phi}_l = (\Phi_{0lm}, \Phi_{1lm}, \ldots, \Phi_{Nlm})
\end{equation}
we perform the contraction
$\vec{\Phi}_l = \mathsf{A}_l \cdot \vec{V}_l$.
At each different point $\vec{r}_1$, we have to re-compute $\vec{V}$
but not $\mathsf{A}$.

The second option is to use the recurrence relation directly
(i.e.~\eqref{eq:rhonl_recurrence} or
\eqref{eq:Phinl_recurrence}). Because we know ahead of time that we
want $N$ iterations of the recurrence relation, we set up the Taylor
series $t(N)$, and use $r_0 + t(N)$ as the dependent variable. The
length of the series then shrinks as we go up the ladder of basis
function evaluations. In practice this second method seems to be
marginally slower than the first one, as more operations on the
abstract Taylor series need to be performed.

\subsection{Unstable modes of a spherical system}
\label{sec:application}

It is important to check whether a basis set constructed according to
the prescriptions of Sec.~\ref{sec:numerical} actually works in
practice. One simple approach might be to just construct $n$ basis
functions, and integrate up the $n \times n$ square of inner products,
testing whether orthogonality is achieved to a given floating-point
precision. However, we know that it is possible to construct basis
sets that are genuinely orthogonal but whose expansions of realistic
mass densities fail to converge in practice, or display other
undesirable numerical effects.\footnote{\refedit{For example, the
    `defective' NFW basis set constructed in
    \citet[Ch.~2]{lilley_thesis}, which does not converge with the
    addition of higher-order angular terms. See also \citet{Sa91}, who
    suggests that ``glitches and generally anomalous behaviour'' in
    the recovery of modes may be related to the form of the chosen
    basis functions -- this should be systematically investigated.}}
Therefore we choose to demonstrate the validity of our approach by
reproducing a physical result from the literature -- the unstable
radial mode of the isochrone model.

We use the discretized Stieltjes method described in
Sec.~\ref{sec:discretized_stieltjes}, where the basis set is adapted
to the isochrone model at zeroth order. However the specific
adaptation is not the crucial part; for this particular application
only the perturbing density needs to be accurately resolved by the
basis elements, so the key feature required of the basis set is only
that it has the correct asymptotic behaviour. To this end, we adapt
the code and method of \citet{Fouvry2022} to show that the same
unstable mode is recovered by our isochrone-adapted basis set. The
part of the code that implements the basis set may be found at
\url{https://github.com/ejlilley/basis}.

\begin{figure}
  \includegraphics[width=0.5\textwidth]{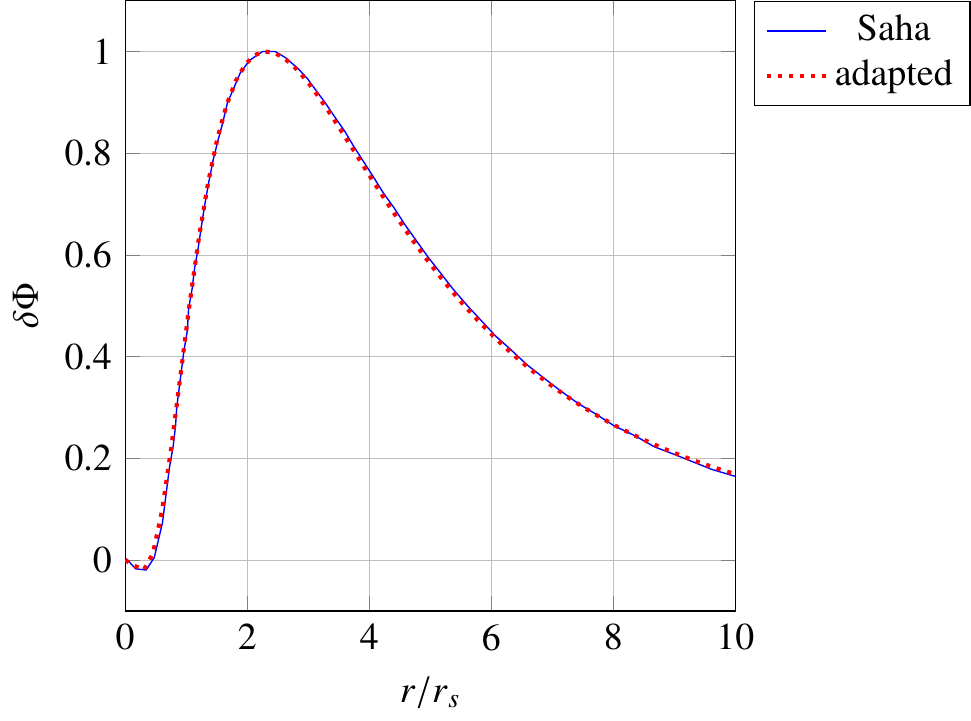}
  \caption{Recovery of the unstable radial mode of the isotropic
    isochrone model. The mode is recovered well despite the low
    ($n_\mathrm{max} = 6$) number of basis functions used.}
  \label{fig:isochrone_mode}
\end{figure}

The details of the computation can be found in
\refedit{\citet{Fouvry2022}}. In brief, we start with knowledge of an
isotropic distribution function that solves the collisionless
Boltzmann equation for the isochrone potential. We also have the
corresponding action and angle coordinates $(\vec{J},\vec{w})$ as a
function of position and momentum, which for the isochrone potential
are known in closed form. Then, each potential basis element must be
Fourier-transformed with respect to the angle coordinates
\begin{equation}\label{eq:angle_fourier}
{\hat \Phi}_{nlm}^\vec{n}(\vec{J}) = (2\pi)^{-3} \int \! d^{\,3}\!\vec{w} \: \expe^{-\imagi \vec{n} \cdot \vec{w}} \: \Phi_{nlm}(\vec{J},\vec{w}),
\end{equation}
out of which a matrix $\mathsf{M}$ is formed\footnote{The azimuthal index $m$
  is set to zero as it does not affect the final result.},
\begin{equation}
(\mathsf{M})_{n_1l_1,n_2l_2} = (2\pi)^3 \sum_{\vec{n}} \int \! d^{\,3}\!\vec{J} \: {\hat \Phi}_{n_1l_10}^\vec{n} \: \overline{{\hat \Phi}_{n_2l_20}^\vec{n}} \: R_{\vec{n}}(s),
\end{equation}
where the $R_{\vec{n}}(s)$ represents the collisionless Boltzmann
operator for a perturbation with growth rate proportional to
$\expe^{st}$. The unstable growing mode then corresponds to a solution
$\vec{A}$ of the matrix equation
\begin{equation}
(\mathsf{M} + \mathbb{I}) \cdot \vec{A} = 0,
\end{equation}
with the vector of coefficients $\vec{A} = \left(A_{nl}\right)$ giving
the expansion of the mode $\delta\Phi$ with respect to the basis $\{\Phi_{nlm}\}$,
\begin{equation}
\delta\Phi = \sum_{nl} A_{nl} \: \Phi_{nl0}.
\end{equation}
\refedit{A plot of this mode is shown in
  Fig.~\ref{fig:isochrone_mode}. The maximum expansion orders were
  $n_\mathrm{max} = 6$ and $l_\mathrm{max} = 2$, with a scale length
  of $\rs = 1$ and a maximum resonance number of
  $n_1^\mathrm{max} = 10$. All our other integration parameters are
  identical to those in \citet[App.~C]{Fouvry2022}, where a matching
  result was obtained using the \citet{CB73} basis set with
  $n_\mathrm{max} = 100$ and $\rs = 20$ -- the mode shape also
  agreeing with the original result of \citet{Sa91}. As mentioned
  previously, it is not strictly necessary to exactly match the
  zeroth-order element of the basis set to the underlying equilibrium
  model. However the basis elements must have the correct asymptotic
  behaviour, so using the isochrone-adapted basis set guarantees that
  this condition is satisfied. Nevertheless, our results do hint that
  accurate mode recovery may be possible with many fewer basis
  elements when the basis is suitably adapted, although we hesitate to
  draw any firm conclusions until a more systematic comparison can be
  drawn.}

Calculating the matrix $\mathsf{M}$ is very computationally expensive,
as it requires multiple truncated infinite summations, over several
indices ($n$, $l$ and the vector of wavenumbers $\vec{n}$). It also
requires two nested integrations, as the Fourier transform
\eqref{eq:angle_fourier} must also be performed numerically. In the
general non-isochrone case a third level of integration is required,
because the action and angle coordinates are no longer known in
closed-form. Any method of reducing this computational effort is
therefore desirable. It is possible that judicious choice of basis
elements and application of their differential-recursion relation
\eqref{eq:Phinl_recurrence} may ameliorate these calculations, but
further investigation is needed.

\section{Discussion and Conclusions}
\label{sec:discussion}

We have reformulated the study of bi-orthogonal basis sets using the
language of Fourier-Mellin transforms. This unexpected development
unifies many previous results into a coherent theoretical
framework. The general idea of generating new potential-density pairs
from old by differentiation is not entirely new. Traditionally this is
accomplished by differentiating with respect to the model's
scalelength -- in particular, \citet{Ao78} found compact expressions
for \citet{CB72}'s \refedit{thin disc} basis by repeatedly applying
the operator $a \partial_a$ (for $a$ the scalelength) and
orthogonalising the resulting sequence of potential-densities by the
Gram-Schmidt process. Subsequently \citet{deZ88}, in the course of
deriving a series of ellipsoidal potential-density pairs, noted that
the operators $r \partial_r$ and $\nabla^2$ obey an important
commutation relation (which we re-derive in
App.~\ref{sec:commutator_proof}). Therefore \citet{Ao78}'s result (and
by extension our algorithm presented here) can be expressed in terms
of the coordinates alone, without reference to an arbitrary
scalelength.

The formalism developed in sections \ref{sec:new}--\ref{sec:theory}
deserves some further interpretation. In particular, the operator
$\dop$ on which the whole development hinges may appear to have been
plucked out of thin air, but it is in fact no accident: $\dop$ is
precisely the infinitesimal generator of the scaling symmetry of the
self-energy inner product \eqref{eq:innerproduct}. To briefly motivate
this, let $S_t$ be a `radial scaling' operator,
\begin{equation}
(S_tf)(\vec{r}) = t^{-5/2} f(t\vec{r}).
\end{equation}
As is immediately evident from dimensional analysis, this preserves
the self-energy, i.e.
\begin{equation}\label{eq:scaling_operator}
\langle S_t f, S_t g \rangle = \langle f, g \rangle.
\end{equation}
The operator $\dop$ is now defined in terms of the infinitesimal
generator of $S_t$,
\begin{equation}
(\dop f)(\vec{r}) \equiv \imagi \left. \frac{d}{dt} (S_tf)(\vec{r}) \right|_{t=0}.
\end{equation}
Differentiating \eqref{eq:scaling_operator} with respect to the
parameter $t$, it is immediately evident that $\dop$ is
self-adjoint\footnote{Multiplication by $\imagi$ in the definition of
  $\dop$ makes it a self-adjoint rather than a skew-symmetric
  operator.}. In Sec.~\ref{sec:fourier_mellin_intro} we implicitly
invoked Stone's theorem from functional analysis to provide a
Fourier-like transform whose integral kernel is the eigenfunction of a
self-adjoint operator. In our case the operator is $\dop$, the
eigenfunction is $\Psi_s$ \eqref{eq:psi_s}, and the resulting integral
transform is exactly the radial part of the Fourier-Mellin transform
that we defined in \eqref{eq:fourier_mellin}. The spherical harmonics
arise from a similar argument applied to the generators of the
coordinate rotations\footnote{The standard construction does not use
  the $\varphi$- and $\vartheta$-generators directly, as they do not
  commute; instead the operators representing the total angular
  momentum $L = \lVert \vec{L} \rVert$ and the $z$-component $L_z$ are
  used. The spherical harmonics are then the joint eigenfunctions of
  $L$ and $L_z$.}.

This line of reasoning suggests that it may be worthwhile to look for
other symmetries of the self-energy inner product, perhaps arising
from other coordinate systems or geometries in which the Laplacian
separates. Given a set of three mutually-commuting operators arising
from three symmetries of the self-energy, we would expect to be able
to construct a basis set formalism similar to that of the present
work. To sketch out what this looks like in full generality, let
$\tau$ be a suitable self-adjoint operator according to the criteria
just described (restricting to one spatial dimension for the sake of
discussion). Then the self-adjointness condition
\eqref{eq:zeta_selfadjoint} combined with the properties of the inner
product \eqref{eq:innerproduct} implies that
\begin{equation}
\nabla^2 \tau = \tau^* \nabla^2,
\end{equation}
where $\tau^*$ is the Hermitian adjoint of $\tau$ with respect to the
ordinary inner product on $L^2$ functions
\eqref{L2_inner_product}. Suppose further that we have found a set of
orthogonal potential functions $\{\Phi_n\}$, with an index-raising
polynomial $p_n(s)$ such that
\begin{equation}
  \Phi_n = p_n(\tau) \Phi_0.
\end{equation}
Then the associated density functions (obeying
$\nabla^2 \Phi_n = \rho_n$) are given by
\begin{equation}
  \rho_n = p_n(\tau^*) \rho_0.
\end{equation}
There are further simplifications involved in Sec.~\ref{sec:theory},
which come about essentially because $\dop = \dop^* + \text{const.}$,
which means that the eigenfunctions of $\dop$ and $\dop^*$ are the
same up to a constant shift in the eigenvalue. Generically we would
expect a different relationship between $\tau$ and $\tau^*$.

The task remaining, which we leave to future efforts, is therefore to
classify the symmetries of the self-energy inner product, in order to
develop expansions that are usefully adapted to different coordinate
systems and geometries. In a sense, the `holy grail' would be the
construction of an expansion adapted to the confocal ellipsoidal
coordinate system, appropriate for studying the equilibrium dynamics
of ellipsoidal galaxies\footnote{Limited work on perturbation analysis
  has been done for the fully ellipsoidal case, including
  e.g.~\citet{Tre76}. There is also some existing work on
  (non-orthogonal) spheroidal basis sets \citep{Ea96,Ro96}.}.

Some symmetries are already known. For example, in Cartesian
coordinates $(x,y,z)$ we trivially have the three cardinal
translations ($x \mapsto x + a$ etc.). Writing down their associated
infinitesimal generators $X = \imagi \partial_x$,
$Y = \imagi \partial_y$ and $Z = \imagi \partial_z$, their joint
eigenfunction $\expe^{\imagi \vec{k} \cdot \vec{r}}$ is just the
kernel of the standard Fourier transform, with the wavevector
$\vec{k}$ taking the role of the (continuous) eigenvalue. The Fourier
transform would therefore play the same role in the resulting basis
set formalism as the Fourier-Mellin transform did in ours
(Sec.~\ref{sec:theory}). Poisson solvers directly using the Fourier
transform are ubiquitous in astrophysical applications, so it would be
interesting to construct a set of `Cartesian' basis functions and
compare their performance with the current state-of-the art.

Other symmetries are known from classical potential theory. Firstly,
the Kelvin transform, which is an inversion in a sphere and preserves
the self-energy up to a sign \citep{Ka76}. However it is not a
continuous symmetry, so there is no associated infinitesimal
self-adjoint operator. Secondly, a symmetry that takes spheres to
concentric ellipsoids (sometimes called homeoids). This maps the
spherical radius to an `ellipsoidal' radius,
$r \mapsto m = \sqrt{x^2/a^2 + y^2/b^2 + z^2/c^2}$. It has long been
known that this transformation preserves the mutual self-energy of any
two charge or mass densities \citep{Ca61}, up to a constant factor
that is essentially just an elliptical integral of the three semi-axes
$(a,b,c)$. We can use this to transform any purely spherical basis
set\footnote{For example, setting $l=m=0$ in any spherical basis set
  considered in this paper.} into one stratified on concentric
ellipsoids. Note however that the \emph{concentric} ellipsoids in this
transformation are distinct from the \emph{confocal} ellipsoids
inherent in the ellipsoidal coordinate system that is more dynamically
relevant due to its relationship to the Stäckel potentials
\citep{deZ85,deZ86}.

Also, we mention some gaps in our analysis. While we purport in this
work to provide a general theory of orthogonal basis sets, there are
some aspects that are still not fully characterised. Firstly, it is
clear from Sec.~\ref{sec:existing} that there exists a connection
between basis sets which have a \emph{classical} index-raising
polynomial $\rawpoly_n(s)$, and those whose potential and density
elements are known in \emph{closed-form} (i.e.~possessing a recurrence
relation independent of $\dop$ or $\dopp$). However, the exact nature
of this connection is unknown, although it is likely related to the
fact that the Hahn-type polynomials appearing in the various
index-raising polynomials obey second-order \emph{difference}
equations\footnote{Contrast the second-order \emph{differential}
  equations obeyed by the polynomials (Gegenbauer etc.) appearing in
  the expressions of many of the known basis sets.}. Secondly, we do
not touch on the issue of basis sets appropriate for finite-radius
systems. This was approached by \citet{Ka76} in the case of
\refedit{thin discs}, using a formalism initially similar to our
own. There are also contributions from \citet{Po81} for finite
spheres, and \citet{Tre76} for finite elliptical discs. In general it
appears to be straightforward to construct basis sets for finite
systems out of polynomials or Bessel functions, but a concrete
connection to our new formalism would be attractive. A more rigorous
form of the argument about completeness in Sec.~\ref{sec:completeness}
would also be desirable, as would a quantitative comparison with basis
sets computed via the Sturm-Liouville approach of \citet{We99}.

Finally, some broader speculation. It is possible that the general
ideas developed here may find applications beyond the solution of
Poisson's equation. In physics we are often required to compute the
inverse of Hermitian operators with a continuous spectrum -- a
well-known example being the Schrödinger operator for certain boundary
conditions and choices of potential. These operators could conceivably
be supplied with a set of (adapted) orthogonal basis functions, by
identifying a suitable commuting set of self-adjoint operators and
then diagonalising their cyclic vectors. Any such basis set then
provides an infinite series representation of the Green's function of
the underlying Hermitian operator\footnote{In the case of the
  Laplacian this is a multipole-like expansion,
  \[\lVert \vec{r} - \vec{r}^\prime \rVert^{-1} = \sum_{nlm}
    \Phi_{nlm}(\vec{r}) \overline{\Phi_{nlm}(\vec{r}^\prime)}.\]}
where the coordinates appear multiplicatively separated in each
term. Such series representations may find use in various
applications. The appearance of tridiagonal Jacobi operators in
particular may presage links to similar numerical methods in quantum
mechanics \citep{Alhaidari2008,Ismail2011}.

\begin{acknowledgements}
  EL and GvdV acknowledge funding from the European Research Council
  (ERC) under the European Union's Horizon 2020 research and
  innovation programme under grant agreement No 724857 (Consolidator
  Grant ArcheoDyn).

  \refedit{We thank Jean-Baptiste Fouvry for granting permission to
    adapt his computer code for the purposes of our
    Sec.~\ref{sec:application}; and also to the referee Michael
    Petersen for numerous helpful suggestions that have strengthened
    the results of this paper.}
\end{acknowledgements}

\bibliographystyle{aa}
\bibliography{paperbib}

\begin{appendix} %

\onecolumn

\section{Self-adjointness of $\dop$}
\label{sec:zeta_selfadjoint}

Let $f,g$ be densities that are non-zero on a $\delta$-dimensional
hyperplane in three-dimensional space ($\delta \leq 3$). Then
\begin{equation}
  \langle f, g \rangle = \doubleint f(\vec{r}) \overline{g(\vec{r}^\prime)}
  \, G(\vec{r}, \vec{r}^\prime),
\end{equation}
where the (three-dimensional Newtonian) Green's function $G$ is
\begin{equation}\label{eq:greens}
  G(\vec{r}, \vec{r}^\prime) = \lVert \vec{r} - \vec{r}^\prime \rVert^{-1}
  = \left(r^2 + r^{\prime 2} - 2rr^\prime \cos{\phi}\right)^{-1/2},
\end{equation}
and $\phi$ is the angle between the two position vectors. Also define
the ordinary $L^2$ inner product,
\begin{equation}\label{L2_inner_product}
(f,g) = \int d^\delta\vec{r} f(\vec{r}) \overline{g(\vec{r})}.
\end{equation}
We write $\theta = r \partial_r$ and
$\theta^\prime = r^\prime \partial_{r^\prime}$. Preliminaries: first
note that
\begin{equation}\label{eq:G_property}
(\theta + \theta^\prime) G = -G,
\end{equation}
and also note that (from integration by parts on $r$)
\begin{equation}\label{eq:L2_property}
(f, \theta g) + (\theta f, g) = -\delta (f,g).
\end{equation}
So we compute
\begin{align}
  \langle f, \theta g \rangle
  &= \doubleint f(\vec{r}) G(\vec{r},\vec{r}^\prime)
   \theta^\prime \, \overline{g(\vec{r}^\prime)} \\ \nonumber
   &= -\doubleint \left[ f(\vec{r}) (\delta G(\vec{r},\vec{r}^\prime)
     \overline{g(\vec{r}^\prime)} + \overline{g(\vec{r}^\prime)}
     \,\theta^\prime G(\vec{r},\vec{r}^\prime) \right] \\ \nonumber
   &= \doubleint\left[ (1-\delta)f(\vec{r}) G(\vec{r},\vec{r}^\prime)
     \overline{g(\vec{r}^\prime)} + f(\vec{r}) \overline{g(\vec{r}^\prime)}
     \,\theta G(\vec{r},\vec{r}^\prime) \right] \\ \nonumber
   &= \doubleint\left[ (1-2\delta)f(\vec{r}) G(\vec{r},\vec{r}^\prime)
     \overline{g(\vec{r}^\prime)} - \overline{g(\vec{r}^\prime)}
     G(\vec{r},\vec{r}^\prime) \,\theta f(\vec{r}) \right] \\ \nonumber
  &= (1-2\delta) \langle f,g \rangle - \langle \theta f, g \rangle,
\end{align}
where to obtain the final result we applied \eqref{eq:L2_property},
then \eqref{eq:G_property}, and then \eqref{eq:L2_property} again. So
define $\dop$ in a $\delta$-dependent way, as
\begin{equation}
  \dop = \imagi (\theta + \delta - 1/2),
\end{equation}
and we can see that
\begin{equation}
  \langle f, \dop g \rangle = \langle \dop f, g \rangle.
\end{equation}
Setting $\delta=3$ (i.e. no restriction to a hyperplane) gives the
appropriate result for spherical geometry. For thin discs we have
$\dopp = \left.\dop\right|_{\delta = 2}$. We could also consider
$\delta = 1$ for an infinite line density.

\section{Commutator of $\dop$ and $\nabla^2$}
\label{sec:commutator_proof}

Working on a $\delta$-dimensional hyperplane again, write the
potential using the Green's function \eqref{eq:greens},
\begin{equation}
  \Phi(\vec{r}) = \singleint \rho(\vec{r}^\prime) G(\vec{r}, \vec{r}^\prime).
\end{equation}
Now apply $\theta$, giving
\begin{align}
  \theta \Phi(\vec{r}) &= \singleint \rho(\vec{r}^\prime)
  \left[ G(\vec{r},\vec{r}^\prime)
    + \theta^\prime G(\vec{r},\vec{r}^\prime)\right] \\ \nonumber
  &= -\Phi(\vec{r}) + \singleint G(\vec{r},\vec{r}^\prime)
  \left( \delta + \theta^\prime \right) \rho(\vec{r}^\prime) \\ \nonumber
  &= (\delta-1) \Phi(\vec{r}) + \singleint G(\vec{r},\vec{r}^\prime)
  \theta^\prime \rho(\vec{r}^\prime),
\end{align}
where we used \eqref{eq:G_property} and then \eqref{eq:L2_property}.
Note that in the spherical ($\delta=3$) case this is equivalent to
\refedit{calculating the following commutator,
\begin{equation}
  \label{eq:commutator}
[\radiallaplacian, \theta] = \radiallaplacian\theta - \theta\radiallaplacian = 2\radiallaplacian,
\end{equation}}
which can be shown directly by differentiation and the Leibniz rule;
however \eqref{eq:commutator} is \refedit{inapplicable to} the thin
disc ($\delta=2$) case, so the previous derivation in terms of the
Green's function is required. Now, writing these results in terms of
the self-adjoint operator $\dop$, we have
\begin{equation}
  \left(\dop +(1-\delta)\imagi\right) \Phi
  = \singleint G(\vec{r},\vec{r}^\prime) \dop^\prime \rho(\vec{r}^\prime).
\end{equation}
Specialising to the spherical case, if for some basis set $\{\rho_n\}$
there exists a suitable index-raising polynomial $\rawpoly_n(s)$, we
have
\begin{equation}
  \rho_n = \rawpoly_{n}(\dop) \rho_0
\end{equation}
for the density functions, and
\begin{equation}
  \Phi_n = \rawpoly_{n}\!\left(\dop -2\imagi\right) \Phi_0
\end{equation}
for the potentials. Analogously, in the \refedit{thin disc} case, we have
\begin{equation}
  \sigma_n = \rawpoly_{n}(\dopp) \sigma_0
\end{equation}
and
\begin{equation}
  \psi_n = \rawpoly_{n}\!\left(\dopp -\imagi\right) \psi_0.
\end{equation}

\section{The Fourier-Mellin transform}
\label{sec:psi_inner_product}

We develop expressions for the forwards and reverse Fourier-Mellin
transform, and the corresponding orthogonality relation. A similar
procedure is followed for both the spherical and the \refedit{thin disc} cases.

\subsection{Spherical case}\label{sec:spherical_fourier_mellin}

We work in spherical polar coordinates $(r, \vartheta, \varphi)$, with
$\vec{r} = r \unitvector$. Our density basis function for the
Fourier-Mellin transform is $\Psi_{slm}$, defined in
\eqref{eq:psi_slm}. The corresponding potential, obeying $\nabla^2
\phi_{slm} = 4 \pi \Psi_{slm}$, is
\begin{equation}
  \phi_{slm}(\vec{r}) = \frac{-4\pi}{K_l(\imagi s)} \: r^{-\imagi s - 1/2}
  \: Y_{lm}(\unitvector),
\end{equation}
where $K_l(\imagi s)$ is defined in \eqref{eq:Kl}. The expansion of an
arbitrary mass density $F$ with respect to the $\Psi_{slm}$-basis is
the Fourier-Mellin transform of $F$:
\begin{align}
  \langle F, \Psi_{slm}\rangle &=
  - \intddd \: F(\vec{r}) \: \overline{\phi_{slm}(\vec{r})} \\ \nonumber
  &= \frac{4\pi}{K_l(\imagi s)} \int_0^\infty r^2 dr \: r^{\imagi s - 1/2}
  \intsphere \: \overline{Y_{lm}(\unitvector)} \: F(\vec{r}) \\ \nonumber
  &= \frac{4\pi}{K_l(\imagi s)} \MellinTransform{F_{lm}(r)}{r}{5/2 + \imagi s},
\end{align}
where
$F_{lm}(r) = \intsphere \: \overline{Y_{lm}(\unitvector)} \:
F(\vec{r})$ are the spherical multipole moments of $F$. Inverting this
using the Mellin inversion theorem \eqref{eq:mellin_inversion}
(choosing the constant $c = 5/2$ in the integral), we have
\begin{equation}\label{eq:fourier_mellin_inversion}
  F(\vec{r}) = \frac{1}{8\pi^2} \sum_{lm} \int_{-\infty}^\infty ds
  \: K_l(\imagi s) \: \Psi_{slm}(\vec{r}) \:
  \langle F, \Psi_{slm} \rangle.
\end{equation}
The potential corresponding to the density $F$ can be expressed
similarly by replacing $\Psi_{slm}(\vec{r})$ in
\eqref{eq:fourier_mellin_inversion} by its potential
$\phi_{slm}(\vec{r})$. Finally, the mutual energy of two densities
$F_1$ and $F_2$ is
\begin{equation}
  \langle F_1, F_2 \rangle = \frac{1}{8\pi^2} \sum_{lm} \int_{-\infty}^\infty ds \:
  K_l(\imagi s) \: \langle F_1, \Psi_{slm} \rangle \: \langle \Psi_{slm}, F_2 \rangle
\end{equation}
and the Fourier-Mellin \refedit{basis functions satisfy} the
orthogonality relation
\begin{equation}
  \langle \Psi_{slm}, \Psi_{t \lambda \mu} \rangle = \frac{8 \pi^2}{K_l(\imagi s)}
  \: \delta_{m\mu} \: \delta_{l\lambda} \: \delta(s-t).
\end{equation}

\subsection{Disc case}
\label{sec:disc_fourier_mellin}

We work in cylindrical polar coordinates $(R, \varphi, z)$, with
$\vec{R} = R\cunitvector$. Define
$\dopp = \imagi (R \partial_R + 3/2)$. Then for two arbitrary
\refedit{thin disc} densities $\sigma_1, \sigma_2 \propto \delta(z)$ we have
$\langle \dopp \sigma_1, \sigma_2 \rangle = \langle \sigma_1, \dopp \sigma_2\rangle$,
i.e. $\dopp$ is self-adjoint (see App.~\ref{sec:zeta_selfadjoint} for
proof, setting $\delta=2$ at the end to give the \refedit{thin disc} case). The
eigenfunctions of $\dopp$ are $\Sigma_s(R) = R^{-\imagi s - 3/2}$ with
real eigenvalue $s$. We then adjoin a cylindrical harmonic to form the
basis functions (Kalnajs' logarithmic spirals)
\begin{equation}
\Sigma_{sm}(\vec{R}) = \Sigma_s(R) \: \expe^{\imagi m \varphi} =  R^{-\imagi s - 3/2} \: \expe^{\imagi m \varphi}.
\end{equation}
Using Toomre's Hankel-transform method we can find the potential
corresponding to this density, which is\footnote{\refedit{Kalnajs
    defines a similar quantity $K(\alpha,m)$, related to our $K_m(s)$
    by $ K_m(\imagi s) = 1/\left(2 K(s,m)\right)$.}}
\begin{equation}
  \psi_{sm}(R, \varphi, 0) =
  \frac{-\pi}{K_m(\imagi s)} \: R^{-\imagi s - 1/2} \: \expe^{\imagi m \varphi},
  \qquad
  K_m(\imagi s) = \left| \frac{\Gamma\!\left(\frac{m+3/2+\imagi s}{2}\right)}
  {\Gamma\!\left(\frac{m+1/2+\imagi s}{2}\right)} \right|^2 \geq 0,
\end{equation}
so that \emph{in the plane} (i.e.~first acting with the full
Laplacian, then afterwards setting $z=0$) we have
\begin{equation}
  \left.\nabla^2 \psi_{sm}(\vec{r})\right|_{z=0} = 4 \pi \Sigma_{sm}(\vec{R}).
\end{equation}
Now we compute the thin disc Fourier-Mellin transform for an arbitrary
thin disc density $\sigma$,
\begin{align}
  \langle \sigma, \Sigma_{sm} \rangle
  &= - \intddd \: \sigma(\vec{R})\: \delta(z) \: \overline{\psi_{sm}(\vec{r})} \\ \nonumber
  &= \frac{\pi}{K_m(\imagi s)} \int_0^\infty R \: dR \: R^{\imagi s - 1/2}
  \int_0^{2\pi} d\phi \: \expe^{-\imagi m \varphi} \sigma(\vec{R}) \\ \nonumber
  &= \frac{\pi}{K_m(\imagi s)} \MellinTransform{\sigma_m(R)}{R}{3/2+\imagi s},
\end{align}
where $\sigma_m(R)$ are the cylindrical multipoles of
$\sigma(\vec{R})$. Using the Mellin inversion theorem to invert this
transform \eqref{eq:mellin_inversion} (with constant $c = 3/2$) gives
\begin{equation}
  \sigma(\vec{R}) = \frac{1}{4\pi^3}\sum_{m=-\infty}^{\infty}\int_{-\infty}^{\infty} ds
  \: K_m(\imagi s) \: \Sigma_{sm}(\vec{R}) \:
  \langle \sigma, \Sigma_{sm} \rangle.
\end{equation}
Therefore the mutual energy of two thin disc densities can be expressed as
\begin{equation}
  \langle \sigma_1, \sigma_2 \rangle = \frac{1}{4\pi^3} \sum_{m=-\infty}^{\infty}
  \int_{-\infty}^{\infty} ds \: K_m(\imagi s) \:
  \langle \sigma_1, \Sigma_{sm} \rangle \: \langle \Sigma_{sm}, \sigma_2 \rangle.
\end{equation}
We also have the orthogonality relation
\begin{equation}
  \langle \Sigma_{sm}, \Sigma_{t\mu} \rangle
  = \frac{4 \pi^3}{K_m(\imagi s)} \: \delta_{m\mu} \: \delta(t-s).
\end{equation}
As noted in Sec.~\ref{sec:disc_polynomials}, these results are
independent of the $z$-dependence of the potential away from the disc
plane\refedit{.}

\section{Orthogonality relation}\label{sec:orth_proof}

\subsection{Spherical case}
\label{sec:spher_orth_proof}

For the inner product of any two density basis functions $\rho_{nlm}$
we have
\begin{align}
  \langle \rho_{nlm}, \rho_{n^\prime l^\prime m^\prime} \rangle
  &= \frac{\delta_{ll^\prime} \delta_{mm^\prime}}{8\pi^2} \int_{-\infty}^\infty ds
  \: K_l(\imagi s) \: \langle \Psi_{slm}, \rho_{nlm} \rangle \: \langle
  \rho_{n^\prime lm}, \Psi_{slm} \rangle \\ \nonumber
  &= \frac{\delta_{ll^\prime} \delta_{mm^\prime}}{8\pi^2} \int_{-\infty}^\infty ds \:
  K_l(\imagi s) \: \langle \Psi_{slm}, \rawpoly_{nl}(\dop) \rho_{0lm} \rangle
  \: \langle \rawpoly_{n^\prime l}(\dop) \rho_{0lm}, \Psi_{slm} \rangle
  \\ \nonumber
  &= \frac{\delta_{ll^\prime} \delta_{mm^\prime}}{8\pi^2} \int_{-\infty}^\infty ds \:
  K_l(\imagi s) \: \overline{\rawpoly_{nl}(s)} \: \rawpoly_{n^\prime l}(s)
  \left|\langle \Psi_{slm}, \rho_{0lm} \rangle\right|^2.
\end{align}
The non-polynomial factors in the above expression are collected into a weight
function $\omega_l(s)$, which can be written explicitly in terms of the Mellin
transform of the zeroth-order density (or a similar expression in
terms of the zeroth-order potential -- see \eqref{eq:omega_l_expressions}),
\begin{align}
  \omega_l(s) &= \frac{K_l(\imagi s)}{8\pi^2} \left|
  \langle \rho_{0lm}, \Psi_{slm} \rangle \right|^2 \\ \nonumber
  &= \frac{2 K_{0l}^2}{K_l(\imagi s)} \left|
  \MellinTransform{\rho_{0l}(r)}{r}{5/2 + \imagi s} \right|^2.
\end{align}
We also assume we have found the (real) monic polynomials orthogonal
with respect to the weight function $\omega_l(s)$, writing them as
$\poly_{nl}(s)$, so that
\begin{equation}
  \int_{-\infty}^{\infty} \omega_l(s) \: \poly_{nl}(s) \: \poly_{n^\prime l}(s)
  = \delta_{n n^\prime} h_{nl}.
\end{equation}
Now write $\rawpoly_{nl}(s)$ in terms of $\poly_{nl}(s)$ as
\begin{equation}
  \rawpoly_{nl}(s) = \imagi^{-n} \: \poly_{nl}(s),
\end{equation}
so that the orthogonality relation for the $\rho_{nlm}$ becomes
\begin{align}
  \langle \rho_{nlm}, \rho_{n^\prime l^\prime m^\prime} \rangle
  &= \delta_{ll^\prime} \delta_{mm^\prime} \int_{-\infty}^\infty ds \: \omega_l(s) \:
  \overline{\rawpoly_{nl}(s)} \: \rawpoly_{n^\prime l}(s) \\ \nonumber
  &= \delta_{ll^\prime} \delta_{mm^\prime} (-\imagi)^{-n} \: \imagi^{-n^\prime} \:
  \int_{-\infty}^{\infty} \omega_l(s) \: \poly_{nl}(s) \: \poly_{n\prime l}(s) \\ \nonumber
  &= \delta_{ll^\prime} \delta_{mm^\prime} \delta_{nn^\prime} h_{nl}.
\end{align}
We have that $\rawpoly_{nl}(\dop)$ is a real operator, because
\begin{align}
  \overline{\rawpoly_{nl}(\dop)}
  &= (-\imagi)^{-n} \: \poly_{nl}(\overline{\dop}) \\ \nonumber
  &= (-\imagi)^{-n} \: \poly_{nl}(-\dop) \\ \nonumber
  &= (-\imagi)^{-n} \: (-1)^n \: \poly_{nl}(\dop) \\ \nonumber
  &= \imagi^{-n} \: \poly_{nl}(\dop) \\ \nonumber
  &= \rawpoly_{nl}(\dop).
\end{align}
This ensures that applying $\rawpoly_{nl}(\dop)$ to a real function
(e.g.~$\rho_{0l}(r)$) gives a real result. Note that we used
$\poly_{nl}(-x) = (-1)^n \poly_{nl}(x)$, which is true for any
orthogonal polynomial where the weight function and domain of
integration are both symmetric.

\subsection{Thin disc case}
\label{sec:disc_orth_proof}

For $\sigma_{nm} = \rawpoly_{nm}(\dopp) \sigma_{0m}$ we have the
orthogonality relation
\begin{align}
 \langle \sigma_{nm}, \sigma_{n^\prime m^\prime} \rangle &=
 \frac{\delta_{m m^\prime}}{4 \pi^3} \int_{-\infty}^\infty ds \: K_m(\imagi s) \:
 \langle \rawpoly_{nm}(\dopp) \sigma_{0m}, \Sigma_{sm} \rangle \:
 \langle \Sigma_{sm}, \rawpoly_{n^\prime m}(\dopp) \sigma_{0m} \rangle \\ \nonumber%
  &= \frac{\delta_{m m^\prime}}{4 \pi^3} \int_{-\infty}^\infty ds \: K_m(\imagi s)\:
  \left| \langle \sigma_{0m}, \Sigma_{sm} \rangle \right|^2 \:
  \rawpoly_{nm}(s) \: \overline{\rawpoly_{n^\prime m}(s)} \\ \nonumber
  &= \delta_{m m^\prime} \int_{-\infty}^\infty ds \: \Omega_m(s) \:
  \rawpoly_{nm}(s) \: \overline{\rawpoly_{n^\prime m}(s)},
\end{align}
where the weight function can be written in terms of either the
zeroth-order potential or density,
\begin{align}
  \Omega_m(s) %
  &= \frac{K_m(\imagi s)}{4 \pi^3} \left| \MellinTransform{\psi_{0m}(R)}{R}{1/2 + \imagi s} \right|^2. \\ \nonumber
  &= \frac{\left| \MellinTransform{\sigma_{0m}(R)}{R}{3/2 + \imagi s} \right|^2}{4 \pi K_m(\imagi s)}.
\end{align}

\section{Classical polynomials}
\label{sec:polys}

Here we record two types of orthogonal polynomial that are used in
Sec.~\ref{sec:existing} -- the \emph{continuous Hahn} and the
\emph{Meixner-Pollaczek} polynomials. We summarise only the properties
that are relevant for our purposes, and direct the reader to other
sources for more comprehensive information \dlmf{18.19}.

These two polynomials are perhaps obscure compared to the well-known
classical polynomials of Jacobi, Laguerre and Hermite. However, a
slight generalisation of the notion of `classical' leads to the Askey
scheme \citep{Koekoek2010}, according to which the continous Hahn and
Meixner-Pollaczek polynomials lie just one level above the Jacobi
polynomials. Like the standard classical polynomials, all Askey
polynomials possess \begin{inparaenum}
\item closed-form expressions in terms of hypergeometric functions,
  and
\item three-term recurrence relations with simple expressions for the recurrence coefficients.
\end{inparaenum}
The latter property means that detailed knowledge about the
polynomials is usually unnecessary, and the end-user can just plug in
the recurrence formulas \eqref{eq:cont_hahn_recur} and
\eqref{eq:meix_poll_recur}.

\subsection{Continuous Hahn}
\label{sec:continuous_hahn}

The continuous Hahn polynomials conventionally take four real
parameters, usually written in terms of two complex parameters:
$(a,b,\overline{a},\overline{b})$. We restrict ourselves to the case
of two real parameters\footnote{Under this parameter restriction these
  polynomials are sometimes referred to as the \textit{continuous
    symmetric Hahn polynomials}.}, so $a = \overline{a}$ and
$b = \overline{b}$, and an explicit representation in terms of a
terminating \hyperg{3}{2} hypergeometric series is
\begin{equation}
\label{eq:continuous_hahn_def}
p_n(s; a, b) = \imagi^n \frac{(2a)_n \: (a + b)_n}{n!}
\hypergeom{3}{2}{-n, n+2a+2b-1, a + \imagi s}{2a, a+b}{1}.
\end{equation}
The orthogonality relation is
\begin{equation}
  \int_{-\infty}^{\infty} p_n(s; a,b) \: p_m(s; a,b) \: ds = \delta_{nm} \: h_n(a,b), \qquad \text{where} \qquad h_n(a,b) = \frac{2\pi \: \Gamma(n+2a) \: \Gamma(n+2b) \Gamma(n+a+b)^2}{n! \: (2n+2a+2b-1) \: \Gamma(n+2a+2b-1)}.
\end{equation}
Note that $p_n(s;a,b)$ is a real-valued polynomial in $s$ of degree
$n$, symmetric in the parameters $a$ and $b$, despite the fact that
$s$ appears (abnormally) in the `parameter' part of the hypergeometric
function. Like any orthogonal polynomial on a symmetric interval, each
individual polynomial is either an even or an odd function, according
to the parity relation $p_n(-s; a, b) = (-1)^n p_n(s; a, b)$. We also
define the monic form of the polynomials,
\begin{equation}
{\hat p}_n(s; a,b) = p_n(s; a,b)/k_n(a,b), \qquad \text{where} \qquad k_n(a,b) = \frac{(n+2a+2b-1)_n}{n!}.
\end{equation}
The monic form obeys the three-term recurrence relation
\begin{align}\label{eq:cont_hahn_recur}
{\hat p}_{-1}(s; a,b) &= 0; \qquad {\hat p}_{0}(s; a,b) = 1; \\ \nonumber
{\hat p}_{n+1}(s; a,b) &= s \: {\hat p}_n(s; a,b) - \beta_n(a,b) \: {\hat p}_{n-1}(s; a,b), \qquad \text{where} \qquad \beta_n(a,b) = \frac{n(n+2a-1)(n+2b-1)(b+2a+2b-2)}{4(2n+2a+2b-3)(2n+2a+2b-1)}.
\end{align}

\subsection{Meixner-Pollaczek}
\label{sec:meixner_pollaczek}

The Meixner-Pollaczek polynomials are another set of orthogonal
polynomials on the interval $(-\infty,\infty)$, depending on two real
parameters $\lambda$ and $\phi$, and have an explicit representation
in terms of a terminating \hyperg{2}{1} hypergeometric function
\begin{equation}
\label{eq:meixner_pollaczek_def}
P_n^{(\lambda)}(x;\phi) = \frac{(2\lambda)_n \: \expe^{\imagi n \phi}}{n!}
\hypergeom{2}{1}{-n,\lambda + \imagi x}{2\lambda}{1 - \expe^{-2\imagi\phi}}.
\end{equation}
The orthogonality relation is
\begin{equation}
  \int_{-\infty}^{\infty} P_n^{(\lambda)}(s; \phi) \: P_m^{(\lambda)}(s; \phi) \: ds = \delta_{nm} \: h_n^{(\lambda)}(\phi), \qquad \text{where} \qquad h_n^{(\lambda)}(\phi) = \frac{2\pi \: \Gamma(n+2\lambda)}{\left(2 \sin\phi\right)^{2\lambda} \: n!}.
\end{equation}
Note that once again the variable $x$ appears in the `parameter' part
of the hypergeometric function. The weight function is
\begin{equation}
  w^{(\lambda)}(x;\phi) =
  \left| \Gamma(\lambda+\imagi x) \right|^2 \expe^{(2\phi-\pi)x}.
\end{equation}
In the case that the parameter $\phi = \pi/2$, the Meixner-Pollaczek
polynomials can be derived from the continuous Hahn
polynomials in two different ways \dlmf{18.21}: if the two parameters
(of the latter) differ by one half
\begin{equation}
p_n(x; a, a+1/2) = \frac{(n+4a)_n}{2^{2n}} P^{(2a)}_n\!\left(2s;\frac{\pi}{2}\right),
\end{equation}
or if the second parameter is taken to infinity
\begin{equation}
\lim_{b \to \infty} \left\{ \frac{p_n(x; a, b)}{(a+b)_n} \right\} = P^{(a)}_n\!\left(s;\frac{\pi}{2}\right).
\end{equation}
The monic form is
\begin{equation}
{\hat P}_n^{(\lambda)}(s; \phi) = P_n^{(\lambda)}(s; \phi)/k_n(\phi), \qquad \text{where} \qquad k_n(\phi) = \frac{\left(2\sin\phi\right)^n}{n!},
\end{equation}
and for the case $\phi = \pi/2$ the three-term recurrence relation is
\begin{align}\label{eq:meix_poll_recur}
  {\hat P}^{(\lambda)}_{-1}(s; \pi/2) &= 0; \qquad {\hat P}^{(\lambda)}_{0}(s; \pi/2) = 1; \\ \nonumber
  {\hat P}^{(\lambda)}_{n+1}(s; \pi/2) &= s \: {\hat P}^{(\lambda)}_n(s; \pi/2) - \beta^{(\lambda)}_n \: {\hat P}^{(\lambda)}_{n-1}(s; \pi/2), \qquad \text{where} \qquad \beta^{(\lambda)}_n = \frac{n(n+2\lambda-1)}{4}.
\end{align}

\section{Exponential disc potential}\label{app:exponential_disc}

We find the potential multipoles corresponding to the exponential disc
density given in \eqref{eq:exp_disc_sigma}, using Toomre's Hankel
transform method as a starting point. Applying the Toomre method to
the disc density gives an auxiliary function
\begin{equation}
  g_m(k) = -2\pi \int_0^\infty R \: \sigma^\text{exp}_m(R) \: J_m(kR) \: dR = -\frac{2^{2+m}\sqrt{\pi}\:\Gamma(m+3/2)\:k^m}{(1 + k^2)^{m+3/2}},
\end{equation}
from which the potential is found via
\begin{equation}\label{eq:sigma_gm_integral}
\psi_m^\text{exp}(R) = \int_0^\infty g_m(k) \: J_m(kR) \: dk.
\end{equation}
Surprisingly, this integral does not appear in the standard tables,
and computer algebra provides an unsatisfactory result involving a
Meijer $G$-function. The $m=0$ case is given \eqref{eq:exp_disc_psi0},
but to derive the higher-orders we need to combine two basic
ideas. Firstly, \citet{LB89} shows how to (in effect) raise the
angular index $m$ of the RHS of \eqref{eq:sigma_gm_integral}, using an
operator (modifying his notation)
\begin{equation}
\Delta_m = R^m \partial_R R^{-m} = \partial_R + (1-m)/R
\end{equation}
that obeys (for generic $\psi_m$ and $g_m$)
\begin{equation}
\Delta_m \psi_m = - \int_0^\infty k \: g_m(k) \: J_{m+1}(kR) \: dk.
\end{equation}
Secondly, inspired by the use of the operator $\theta = R \partial_R$
in the main part of the present work, we apply it to
\eqref{eq:sigma_gm_integral} and perform some integration by parts to
find (writing $\theta_k = k \partial_k$)
\begin{equation}
(\theta + 1) \psi_m = - \int_0^\infty \theta_k\!\left(g_m(k)\right) J_{m}(kR) \: dk.
\end{equation}
It remains to apply linear combinations of $\theta$ and $\Delta_m$ to
\eqref{eq:sigma_gm_integral}, and then rearrange the terms inside the
integral sign according to our knowledge of $g_m(k)$ such that
\refedit{only a term proportional to $g_{m+1}(k) \: J_{m+1}(kR)$}
remains on the RHS. The result is the recursion relation given in
\eqref{eq:exp_disc_psi}.

\section{Exact moments for the isochrone}
\label{sec:isochrone_exact}

Using the expressions for the isochrone model in \eqref{eq:isochrone},
we seek the modified moments of self-energy
\begin{equation}
{\tilde \mu}_{jl} = \left\langle {\tilde p}_{jl}(\dop) \rho_{0l}^\text{iso}, \rho_{0l}^\text{iso} \right\rangle = \int_{-\infty}^{\infty} ds \: \omega_l^\text{iso}(s) \: {\tilde p}_{jl}(s) = -\!\!\int_0^\infty\!\! dr \: r^2 \rho_{0l}^\text{iso} \: {\tilde p}_{jl}\!\left(\dop - 2\imagi\right) \Phi_{0l}^\text{iso}.
\end{equation}
The auxiliary polynomials ${\tilde p}_{jl}(s)$ here are the monic
Hermite polynomials\footnote{The choice of auxiliary polynomial
  \emph{does} affect the values of the modified moments
  ${\tilde\mu}_{jl}$; but in principle it does \emph{not} affect the
  final value of the recurrence coefficients $\beta_{jl}$, other than
  indirectly via its effect on the numerical stability of the
  algorithm.}. To facilitate variable substitutions in this integral,
it is useful to rewrite both $\Phi_{0l}^\text{iso}$ and
$\rho_{0l}^\text{iso}$ in rationalised-surd form,
\begin{align}
  \Phi_{0l}^\text{iso}(r) &= -\frac{\left(1 - \sqrt{1 + r^2}\right)^{1+2l}}{r^{2+3l}}, \\ \nonumber
  \rho_{0l}^\text{iso}(r) &= -\frac{(1+2l)\left(1 - \sqrt{1 + r^2}\right)^{2+2l}}{4\pi r^{4+3l} \left(1+r^2\right)^{3/2}} \left[1 + 2(1+l)\sqrt{1+r^2}\right].
\end{align}
We also define an auxiliary quantity $K_{jl}$,
\begin{equation}
K_{jl} = (1+2l) \int_0^1 dt \: (1+t)^{-5/2-3l} (1-t)^{l+1/2} t^{1+2l+j} = \frac{(1+2l) \Beta(l+3/2,2l+j+2)}{2^{3l+5/2}} \hypergeom{2}{1}{l+3/2,3l+5/2}{3l+j+7/2}{\frac{1}{2}}.
\end{equation}
In fact $K_{jl}$ can always be reduced (by a computer algebra system
such as \textsc{Mathematica}) to a form $a + b/\pi$ with $a,b$
rational. Writing the integral for the zeroth-order self-energy
${\mu}_{0l}$ with the variable substitution
$t = 1/\sqrt{1+r^2}$, we find that
\begin{align}
{\mu}_{0l} &= -\!\!\int_0^\infty\!\! dr \: r^2 \rho_{0l} \: \Phi_{0l} \\ \nonumber
                  &= (1+2l) \int_0^1 dt \: t^{1+2l} (1-t)^{1/2+l} (1+t)^{-5/2-3l} \left[t + 2(l+1)\right] \\ \nonumber
                  &= K_{1l} + 2(l+1)K_{0l}. %
\end{align}
To find the higher-order moments, consider the following polynomial
$Q_{jl}(t)$ of degree $2j-1$,
\begin{equation}\label{eq:isochrone_Qjl}
Q_{jl}(t) = \left[\Phi_{0l}^\text{iso}(r(t))\right]^{-1} \: {\tilde p}_{jl}(\dop - 2\imagi) \: \Phi_{0l}^\text{iso}(r(t)),
\end{equation}
and note that $r\partial_r = -t(1-t^2)\partial_t$. Using the
recurrence relation \eqref{eq:ptkl_recurrence} for the auxiliary
polynomials ${\tilde p}_{jl}(s)$ we can therefore write $Q_{jl}(t)$
recursively as
\begin{align}
  Q_{0l}(t) &= 1, \\ \nonumber
  Q_{1l}(t) &= \frac{\imagi}{2}(1+2l)(2t-1), \\ \nonumber
  Q_{jl}(t) &= \left[Q_{1l}(t) - \imagi t(1-t^2)\partial_t\right] Q_{j-1,l}(t) - {\tilde \beta}_{j-1,l} Q_{j-2,l}(t).
\end{align}
Writing out the polynomial explicitly as
$Q_{jl}(t) = \sum_{k=0}^{2j-1} q_{jkl} t^k$, we have the following
recurrence on the coefficients $q_{jkl}$,
\begin{align}
  q_{jkl} &= 0 \text{  when  } k<0 \text{  or  } k>2j-1, \\ \nonumber
  q_{00l} &= 1, \\ \nonumber
  q_{10l} &= \frac{-\imagi}{2}(1+2l), \\ \nonumber
  q_{11l} &= \imagi(1+2l), \\ \nonumber
  q_{jkl} &= \imagi\left[(1+2l)q_{j-1,k-1,l} - (1/2+l+k)q_{j-1,kl} + (k-2)q_{j-1,k-2,l}\right] - {\tilde \beta}_{j-1} q_{j-2,kl}.
\end{align}
Now insert this into the integral for ${\tilde \mu}_{jl}$, finally
giving us the modified moments
\begin{equation}
{\tilde \mu}_{jl} = \sum_{k=0}^{2j-1} q_{jkl} \left( K_{k+1,l} + 2(l+1)K_{kl}\right).
\end{equation}

\subsection{Specific exact coefficient expressions}

Plugging the expression for the modified moments into the modified
Chebyshev method described in Sec.~\ref{sec:modified_chebyshev}, we
can get exact expressions for the recurrence coefficients
$\beta_{nl}$. Setting
$b_l = B_{1/2}\!\left(3l+\frac{5}{2},-l-\frac{1}{2}\right)$ (for
$\Beta_z\!(a,b)$ an incomplete Beta function), the first few are

{\footnotesize
\begin{align*}
  \beta_{0l} &=    \frac{(2 l+1)^2 (2 l)! \Gamma \left(l+\frac{1}{2}\right) \left(4^{-l}-2 (2l+1) b_l\right)}{24\pi  \Gamma \left(3 l+\frac{3}{2}\right)} \\
  \beta_{1l} &=    \frac{(2 l+1) (2 l+3) \left(4 l-2^{2 l+1} (2 l+1) (8 l+5)b_l+1\right)}{4^{l+2} (2 l+1)b_l-8} \\
  \beta_{2l} &=  -\frac{4 l (3 l (20 l+39)+67)+4^{l+1} (2 l+1)^2 b_l \left(4 l (l (8 l-23)-58)-4^l (2 l-1)(2 l (16 l (4 l+9)+95)+35) b_l-93\right)+55}{16 \left(4 l+4^{l+1} (2l+1)^2 b_l\left(4^l (8 l+5) b_l-3\right)+1\right)} \\
\end{align*}}

\end{appendix}

\end{document}